# A more comprehensive habitable zone for finding life on other planets

**Ramses M. Ramirez** [1*]

[1] Earth-Life Science Institute, Tokyo Institute of Technology; rramirez@elsi.jp
* Correspondence: rramirez@elsi.jp; Tel.: +03-5734-2183

**Abstract:** The habitable zone (HZ) is the circular region around a star(s) where standing bodies of water could exist on the surface of a rocky planet. Space missions employ the HZ to select promising targets for follow-up habitability assessment. The classical HZ definition assumes that the most important greenhouse gases for habitable planets orbiting main-sequence stars are $CO_2$ and $H_2O$. Although the classical HZ is an effective navigational tool, recent HZ formulations demonstrate that it cannot thoroughly capture the diversity of habitable exoplanets. Here, I review the planetary and stellar processes considered in both classical and newer HZ formulations. Supplementing the classical HZ with additional considerations from these newer formulations improves our capability to filter out worlds that are unlikely to host life. Such improved HZ tools will be necessary for current and upcoming missions aiming to detect and characterize potentially habitable exoplanets.

## 1. Introduction

The habitable zone (HZ) is the circular region around a star (or multiple stars) where standing bodies of liquid water could exist on the surface of a rocky planet (e.g., [1,2]). Principally, the HZ is a navigational tool used by space missions to select promising planetary targets for follow-up observations. Although a planet located within the HZ is not necessarily inhabited, and additional information would be required to make such a determination, it remains the most useful roadmap for targeting potentially habitable worlds.

It is possible to invent alternate HZ definitions for solvents other than water [3]. Perhaps life thrives within the methane seas of Titan [4–6] and indeed this author hopes that a mission equipped to answer this question will visit the Saturnian moon someday. In contrast to the planets in our solar system, exoplanets cannot be visited so readily. Titan exo-analogues would be too far away from their stars and too faint for signs of life to be inferred with present day technology. The spectral signatures that can be expected from such worlds are also unknown. Therefore, the search for extraterrestrial life on exoplanets must be *presently* limited to a search for liquid water until future technological and biochemical breakthroughs can allow for a detailed exploration of planetary surfaces capable of hosting other solvents.

As suggested recently [2], the addition of the phrase "standing bodies of water" to the HZ definition emphasizes that habitable zone planets incapable of supporting more than seasonal trickles of surface water, like Mars , should not be considered for follow-up astrobiological observations. This is because any life residing within such small amounts of water would be unlikely to modify the atmosphere in a detectable manner. To be clear, such a search does not exclude potential 'Dune planets', or desert worlds that exhibit water inventories smaller than the Earth [7,8], so long as sufficiently large bodies of water near the surface capable of modifying atmospheric spectra in a detectable way are present. Likewise, although moons like Europa and Enceladus contain large seas, these water bodies are capped by global ice layers that preclude any significant interactions between potential subsurface biota and the atmosphere. Thus, the HZ must also be defined for planets in which the atmosphere and surface water bodies are in direct contact.

It is also assumed that life requires a stable liquid or solid surface over which it can evolve. For this reason, it is not presently thought that gas giant atmospheres [9,10] present likely abodes for life. Chaotic atmospheric motions and complex chemical reactions may thwart life's emergence. Until convincing evidence to the contrary arises, all discussion here assumes that rocky planets are the likely locales for alien life.

Even with such restrictions, the HZ can target a wide variety of worlds potentially consisting of life both like and unlike that on this planet. For example, in regions of higher pressure, the highest temperature sustainable by Earth life is ~ 394 K [11] while the highest possible theoretical limit is ~453 K [12]. Both of these numbers are considerably lower than the temperature corresponding to the critical point of water (647 K), which is the highest possible temperature for surface water to exist in liquid form on a rocky planet with an Earth-like inventory [1,13]. The significant versatility of this liquid water HZ will be further demonstrated in this work.

Here, I review the latest advances in planetary habitability studies, especially those related to the habitable zone, and discuss how the concept has evolved in recent years. I also discuss planetary habitability within the context of our solar system, which provides context for understanding exoplanetary habitability. This review will primarily focus on the *circumstellar* habitable zone, although habitable zones in binary star systems will also be discussed. However, I will not delve into related concepts derived at different spatial scales, such as the galactic habitable zone [14,15], local exomoon habitable zones around giant planets [16], or habitable zones for subterranean biospheres [17]. Although these concepts should be explored in the future, upcoming technologies have the best chance of detecting extraterrestrial life in the circumstellar habitable zone and (possibly) habitable planets in binary systems. Even given current technological limitations, I show that this HZ is still rather inclusive, capable of finding an amazing variety of extraterrestrial life, including that which is unlike what is presently on this planet. I discuss how alternative and classical HZ definitions can complement one another and maximize our chances of successfully finding extraterrestrial life.

## 2. The classical habitable zone

This section summarizes the basic theory behind the classical definition.

*2.1 Additional assumptions*

The classical HZ definition of Kasting et al. [1] remains the most popular incarnation of this navigational tool, which includes additional assumptions that complement those discussed in the previous section. The classical HZ posits that $CO_2$ and $H_2O$ are the most important greenhouse gases for habitable planets throughout the universe. This is based on the idea that a long-term (i.e., over geological timescales) carbonate-silicate cycle maintains the habitability of the Earth, as well as other potentially habitable planets, by regulating the transfer of carbon between the atmosphere, surface, and interior [18].

The classical HZ is concerned with the detection of both simple and complex life. "Complex life" includes advanced lifeforms like animals, higher plants, and (possibly) even intelligence. Kasting et al. [1] explicitly mention that the HZ is designed to also find habitable planets that may be unsuitable for humans . In addition, Kasting et al. [1] speculate that intelligent life may take at least 1 or 2 billion years to develop (Kasting et al.[1]), arguing that advanced lifeforms may be found on habitable planets orbiting F – M stars (e.g., [1]). As I discuss in Section 6, however, given the theoretical (e.g., atmospheric, geological, and biological) advances in recent years, an even wider spectral range should be employed if we wish to maximize our chances of finding life elsewhere.

Likewise, the classical HZ definition is concerned with carbon-based life. For the liquid water HZ, this assumption is probably reasonable because silicon-based compounds, particularly silanes, although soluble, can only exhibit the flexibility of carbon-based compounds in very cold environments, like those on Triton or Titan [19]. Indeed, silicon reactions can be catalyzed in such cold environments even at low sunlight levels, and its versatility has been demonstrated in engineering applications [20]. However, silicon-based compounds are likely to be less effective in the warmer environments characterized by surface liquid water (ibid). Plus, all known life requires liquid water to survive. For these reasons, the focus on carbon-based life, which presumes liquid water, is probably reasonable for the near future.

*2.2 Effective Stellar Flux*

The effective stellar flux ($S_{EFF}$) is a key quantity used in HZ calculations and is defined as the normalized flux required to maintain a given surface temperature (e.g., [1]). It is an expression of energy balance which balances the net incoming solar radiation ($F_s$) and the net outgoing radiation ($F_{IR}$), both of which are calculated at the top of the atmosphere (TOA). This leads to the common definition [1] (equation 1):

$$S_{EFF} = \frac{F_{IR}}{F_s} = \frac{S}{S_o} \ , \tag{1}$$

Here, $S$ is the flux received by the planet whereas $S_o$ is the normalized solar flux received by the Earth (~1360 W/m$^2$). In radiative equilibrium, $F_{IR}$ must equal $F_s$ and $S_{EFF}$ is equal to 1. Thus, values of $S_{EFF}$ that differ from unity measure how far a planetary atmosphere deviates from radiative equilibrium, which is also equivalent to deviations in the normalized flux ($S/S_o$). Radiative equilibrium is imposed for the Earth located at 1 AU but not for the inner and outer edges of the HZ. However, for HZ calculations it suffices only to know how much more (or less) stellar energy a planet receives with respect to Earth, after also accounting for the changes in planetary albedo ($A_p$) through energy balance (equation 2):

$$\frac{S_o}{4}(1 - A_p) = \sigma T^4 = F_s, \tag{2}$$

Here, $T$ is surface temperature. Thus, $S_{EFF}$ values greater than 1 also represent smaller semi-major axis distances, corresponding to distances of increased stellar (or solar) flux whereas small values correspond to farther distances of reduced stellar (or solar) flux. The inverse-square law can be rearranged to solve for the distance required to support a given value of $S_{EFF}$ (equation 3):

$$d = \sqrt{\frac{L/L_{sun}}{S_{EFF}}} \ , \tag{3}$$

Where $L/L_{sun}$ is the stellar luminosity in solar units and $d$ is the orbital distance (in AU). Thus, assigning $L/L_{sun} = S_{EFF} = 1$ yields a value of $d = 1$ AU for the Earth. Equation 3 can also be used to compute the effective flux minus the albedo effect. The flux received by the Earth before the energy enters the atmosphere is ~1360 W/m² [1]. At Mars' distance of 1.52 AU, equation 3 predicts that $S_{EFF}$ is ~0.43, which is equal to a flux of (0.43 × 1360) ~585 W/m².

*2.2 The inner edges of the classical habitable zone*

At distances that are close enough to the star, planetary temperatures rise, and photolysis is enhanced, which is a process in which upper atmospheric H₂O vapor molecules dissociate into H⁺ and OH⁻ ions. Although often misrepresented in the literature, the inner edge boundaries are not calculated using the boiling point of water. Indeed, the inner edge has nothing to do with the boiling point of water. On a planet with an Earth-like non-condensable inventory ( ~300 – 400 ppm $CO_2$), a moist greenhouse (a pessimistic inner edge distance) can be triggered when the upper atmospheric water vapor mixing ratio above the cold trap exceeds ~2x10⁻³, which corresponds to a computed mean surface temperature of 340 K [1]. Above the cold trap the air is too cold for water vapor to convect to even higher levels. However, at this surface temperature, water vapor photolysis above the cold trap becomes efficient enough to remove an Earth-like surface water inventory via escape to space within the age of the solar system (or 4.5 billion years)(Figure 1). An example calculation is shown below.

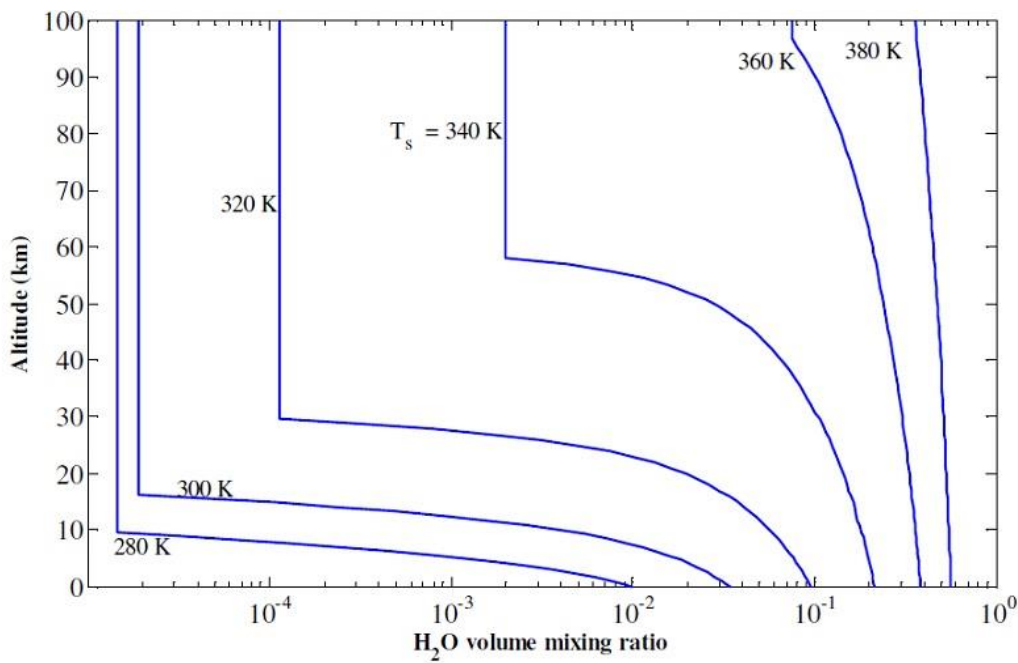

**Figure 1:** Vertical water vapor (volume) mixing ratio profiles for the Earth at different mean surface temperatures. An abrupt increase in the water vapor mixing ratio occurs at ~340 K, after which point the water vapor surface mixing ratio is ~20% [21] and the moist greenhouse is triggered (Adapted from Ramirez [22]). The cold trap is often coincident with the tropopause, which separates the upper atmosphere from the convective region, "trapping" water vapor to lower atmospheric levels (i.e., below the tropopause).

If H escape is assumed to be limited by its diffusion through the homopause, which is the fastest possible escape rate per unit area ($\phi$ [cm⁻²s⁻¹]) at these concentrations, then:

$$\phi = \frac{b}{H_a} fH_2, \quad (4)$$

Here, $fH_2$ is the total H₂ mixing ratio at the homopause (~2x10⁻³ for the moist greenhouse), $H_a$ is the atmospheric scale height, and *b* is a constant the describes how H₂ diffuses through a background atmosphere. For Earth, the quantity $b/H_a$ is ~$3.3 \times 10^{13}$ given $H_a = 7.5 \times 10^5$ cm and assuming $b = 2.5 \times 10^{19}$ cm$^{-2}$s$^{-1}$ [23]. This yields an escape rate per unit area of ~$6.6 \times 10^{10}$ molecules/cm$^2$/sec. The Earth's surface area is ~$5.1 \times 10^{18}$ cm$^2$, resulting in an H escape rate of ~$3.4 \times 10^{29}$ molecules/sec. The mass of $H_2$ in Earth's oceans is ~$(1/9) \times 1.4 \times 10^{21}$ kg or $1.56 \times 10^{20}$ kg (~$4.7 \times 10^{46}$ molecules). Thus, it takes ~ ($4.7 \times 10^{46}/3.4 \times 10^{29}$) $1.4 \times 10^{17}$ seconds (~4.5 billion years) for all of the H molecules in Earth's oceans to escape to space, as mentioned above.

Planetary atmospheres with even higher non-condensible inventories than the Earth are diluted with respect to their water vapor concentrations, requiring that the moist greenhouse be triggered at even higher mean surface temperatures. The opposite is true for worlds with lower non-condensible inventories [1,24].

A more optimistic inner edge is located even closer to the star, commencing when the net absorbed stellar flux exceeds the net outgoing radiation at the top of the atmosphere, which triggers a rapid and uncontrollable runaway greenhouse that can desiccate the planet on shorter (~thousand to million year) timescales. In our solar system, this "runaway greenhouse limit" occurs at ~ 0.95 AU according to models (e.g., [25]). Previous work had suggested that the runaway greenhouse on a planet with an Earth-like surface water inventory occurs at the critical point for water (i.e., mean surface temperature of 647 K and surface pressure above ~220 bars) [1,26]. However, recent calculations find that once the net absorbed solar flux exceeds the thermal infrared flux, radiative energy balance is no longer possible and the runaway greenhouse gets triggered at temperatures well below 400 K (e.g., [13,25,27]). Such temperatures are potentially consistent with suggested upper limits for life on Earth [11]. Again, on a planet with a smaller water inventory than our planet, the runaway greenhouse can be triggered at even lower surface temperatures (e.g., [1]). This is because less solar energy would be needed to devolatilize the surface.

Nevertheless, the exact conditions under which a moist greenhouse may occur is currently under debate and further modeling will be needed. Leconte et al. [25] had found that the moist greenhouse is bypassed in atmospheres that become steadily warmer, immediately transitioning to the runaway greenhouse state instead (see below). In contrast, subsequent calculations find a moist greenhouse state before initiation of the full runaway [28,29].

Some 3-D models using similar convection schemes find that the moist greenhouse may be triggered at lower mean surface temperatures for M-stars [27,30]. However, these results are in conflict with other 3-D and 1-D models that do not find such behavior [25,29]. More modeling is needed to resolve this discrepancy although it may be related to differences in stratospheric temperatures computed or assumed [31].

*2.3 The carbonate-silicate cycle and the outer edges of the classical habitable zone*

A major difference between Earth and the remaining solar system planets is that our planet exhibits plate tectonics (e.g., [32]). Plate tectonics is thought to regulate the carbonate-silicate cycle, which maintains habitability on the Earth over long billion year timescales [33]. This cycle is essentially an interplay between volcanism and silicate weathering processes, allowing carbon to be efficiently recycled between the atmosphere, surface, and interior on Earth, and also perhaps on potentially habitable exoplanets (e.g., [1,28]). Thus, worlds with inefficient volcanism, such as stagnant-lid super-Earths, would not fall under this category [35]. If it weren't for this cycle, it is thought that the HZ would be much narrower, with Earth's surface freezing if it moved out past ~1.05 AU (e.g., [7,36]).

On a planet with a carbonate-silicate cycle, times of increased volcanism would release higher concentrations of greenhouse gases into the atmosphere, including $CO_2$. However, rainfall intensifies as temperatures rise , leading to an increased production of carbonic acid which intensifies surface weathering reactions, resulting in a decrease in atmospheric $CO_2$ until the atmosphere achieves a new equilibrium [18]. Thus, the carbonate-silicate cycle acts as a planetary thermostat that regulates mean surface temperatures [33,37].

This theory predicts that $CO_2$ concentrations on habitable planets should increase at higher semi-major axis distances. This is because stellar fluxes decrease farther away, which reduces both surface temperatures and rainfall. In response, volcanically outgassed $CO_2$ will accumulate until a new atmospheric equilibrium is achieved at the higher mean surface temperature. However, at distances far enough away from the star, atmospheric $CO_2$ pressures become so high that $CO_2$ itself begins to condense out of the atmosphere. The combination of this, and the increased reflectivity at these high $CO_2$ levels, reduces the greenhouse effect. At distances beyond the outer edge of the classical $CO_2$-$H_2O$ HZ, the combination of these two cooling mechanisms outstrips the greenhouse effect and the planet becomes too cold to support standing bodies of liquid water (e.g., [1,26]).

In our solar system, this $CO_2$ maximum greenhouse limit occurs at ~1.67 AU at a $CO_2$ pressure of ~ 8 bars (Figure 2). In contrast, HZ planets orbiting cooler stars exhibit reduced Rayleigh scattering and increased near-infrared absorption, which delays the maximum greenhouse limit to even higher $CO_2$ pressures. For instance, the maximum greenhouse limit for an M8 star occurs at a $CO_2$ pressure of ~ 20 bars (Figure 2).

An alternative outer edge limit had been originally defined at an orbital distance beyond which $CO_2$ first condenses (the so-called fist $CO_2$ condensation limit)[1]. At the time it was thought that $CO_2$ clouds may cool planets. However, this limit had been abandoned as subsequent work found that $CO_2$ clouds generally warm planets, even if the warming is not very much [38–40].

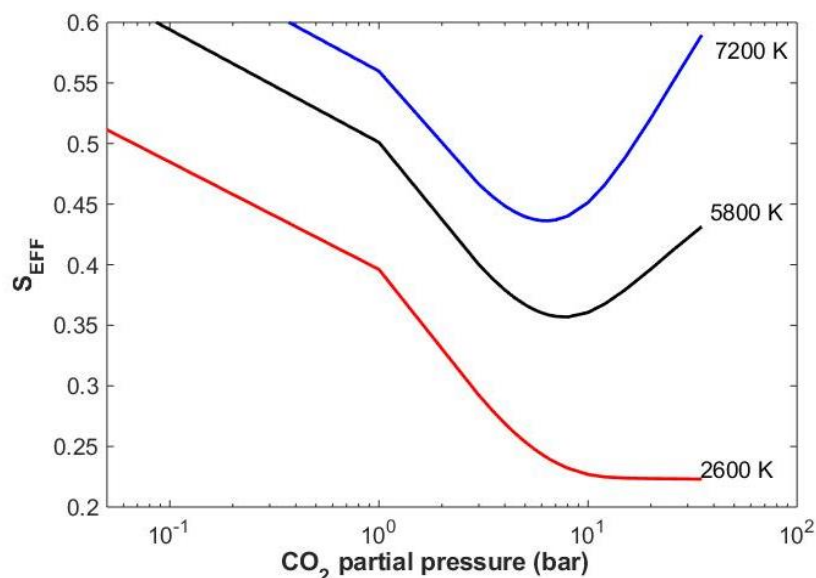

**Figure 2:** Plot of the effective stellar flux ($S_{EFF}$) vs. $CO_2$ partial pressure for F0 (7200 K, blue), G2 (5800 K, black) and M8 (2600 K, red) stellar types. $S_{EFF}$ decreases from the right to the curve minima

because stellar radiation decreases farther away from the Sun, requiring a stronger greenhouse effect (higher $CO_2$ pressures via the carbonate-silicate cycle) to maintain habitable surface temperatures. $S_{EFF}$ increases beyond the curve minimum because the planet cools at the larger distances associated with increased Rayleigh scattering and $CO_2$ condensation. Thus, the maximum greenhouse effect of $CO_2$ occurs at the curve minima (Based on calculations from Kasting et al. [1] and Kopparapu et al.[26]).

Such a carbonate-silicate cycle may help explain the faint young Sun paradox, which is the perceived contradiction between abundant evidence for warm conditions on the early Earth (~2 – 3.9 Ga ) in spite of a fainter young Sun [41]. It was originally thought that $CO_2$ levels were not high enough (< 25 times that of today's) to warm the planet at ~ 2.5 Ga according to earlier reconstructions of paleosoil data [42]. At these later times, additional explanations (e.g., lower cloud cover, fractal hazes, higher $CH_4$ and/or $H_2$ concentrations) may have been necessary to completely resolve the paradox (e.g., [43–46]). However, updated estimates indicate that these $CO_2$ levels could have been underestimated by about an order of magnitude, suggesting that higher $pCO_2$ may have been enough to resolve the paradox without invoking additional mechanisms [47] Plus, very high atmospheric $CO_2$ pressures could have resolved the paradox at the earliest times (~3.9 Ga) as well [45].

*2.4 The classical habitable zone with empirical limits*

Alternatively, complementary HZ limits can be derived that are not based on models, but on empirical observations of our solar system. For example, the inner edge of this *empirical* HZ is defined by the stellar flux received by Venus when we can exclude the possibility that it had standing bodies of water on its surface (~ 1 Ga ; [1]). That is, if Venus had surface fluvial features suggestive of a once habitable planet, they have been absent for at least 1 billion years given that the present surface is free of such features. The flux received at Venus's orbit today is ~ 1.92 times that received by the Earth. According to solar evolution models [48], the Sun was only 92% as bright ~1 Ga while Venus then had received (0.92 × 1.92) ~1.77 times the energy received by the Earth today. Thus, the corresponding Recent Venus limit today is at ~0.75 AU according to equation 3.

Likewise, a similar empirical limit can be defined for the outer edge. This "early Mars limit" is based on geological evidence that Mars may have been a habitable planet ~3.8 Ga, which occurred when the Sun was ~75% as bright as today. The flux received by present Mars is ~43% that of Earth's according to the inverse square law (and equation 3). Thus, $S_{EFF}$ in this case is ~0.32 (0.75 × 0.43), yielding an early Mars limit distance ($L/L_{sun}$ = 1) of ~ 1.77 AU according to equation 3. These conservative and empirical HZ limits for F – M stars can be combined into a graphical depiction of the well-known classical HZ (Figure 3).

As seen in Figure 3, whereas the differences between the conservative (maximum greenhouse) and optimistic (early Mars) outer edge limits are relatively small, those for the inner edge are much larger. This is because the early Mars limit is supported by geologic evidence whereas the Recent Venus limit is simply inferred from an absence of data and so the corresponding uncertainties regarding the location of the classical inner edge are much larger.

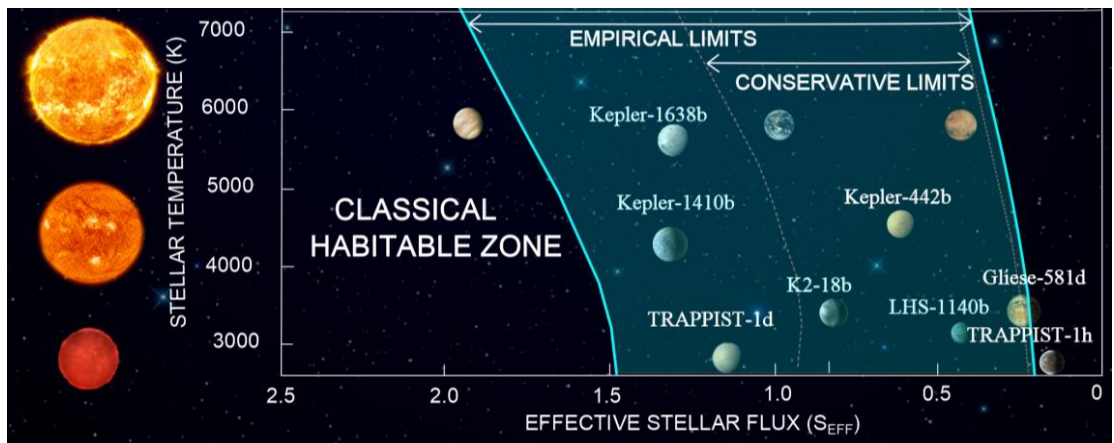

**Figure 3**: The classic $CO_2$-$H_2O$ habitable zone with stellar effective temperature ($T_{EFF}$ = 2,600 – 7,200 K) as a function of the effective stellar flux ($S_{EFF}$). From left to right, the "Recent Venus" and "early Mars" limits (solid blue curves) are the empirical (optimistic) classical HZ limits whereas the "Leconte et al" and "Maximum Greenhouse" limits compose the conservative (pessimistic) classical HZ limits. Some solar system planets and confirmed exoplanets are shown. Based on work from Kopparapu et al. [26,49].

## 3. Habitability case studies: Venus and Mars

The subsequent sections delve into the specific cases of Mars and Venus, which are examples of planets that have significantly diverged from our own, providing the rationale for the classical HZ as defined [1].

*3.1 The fate of Venus*

In conjunction with HZ theory, the carbonate-silicate cycle could potentially explain the state of the current Venusian atmosphere. The inner edge of the HZ has likely been past Venus' orbital distance (~0.72 AU) for at least the past several hundred million years, possibly 1 billion years [1]. Thus, Venus could have been in a moist or runaway greenhouse state for a comparable (if not longer) amount of time, which would have depleted any hypothetical surface ocean. According to this idea, as the planet desiccates, weathering reactions slow down and the water incorporated within subducting plates decreases. This causes the plates to become too brittle for subduction and plate tectonics ceases. The resultant cessation in silicate weathering then leads to the buildup of the currently observed ~90-bar $CO_2$ atmosphere. Alternatively, perhaps Venus never had a surface ocean, as the water was lost to a runaway greenhouse during the accretion process itself [50,51]. The high atmospheric escape rates in both scenarios are consistent with the very high measured atmospheric D/H ratio (>100x greater than that for the Earth; [52]). However, the disappearance of the resultant atmospheric oxygen is difficult to explain if Venus had managed to acquire a surface ocean. A leading idea is that a magma ocean had formed during accretion, removing the remnant atmospheric oxygen via drawdown [50]. I examine the efficiency of this oxygen drawdown with the following calculation.

Magma oceans can be of many different sizes and depths, including global bodies that are thousands of km deep [53]. If I assume a ~2000 km deep global ocean on Venus (radius = 6052 km), its volume would be ~ $6.5 \times 10^{11}$ km$^3$, after subtracting out the core and solid mantle volumes. Assuming a typical mantle density (4000 kg/m$^3$ from ref. [54]), the magma ocean mass is ~$2.6 \times 10^{24}$ kg. The degree of magma ocean drawdown is obtained by calculating the FeO that can be oxidized to FeO$^{1.5}$ [55]. I assume Fe$^{3+}$/(Fe$^{2+}$+Fe$^{3+}$) ratios and Fe$^{2+}$ (by weight) values consistent with the Earth (0.025 and 8%, respectively; ibid), which yields ~0.21% Fe$^{3+}$ or ~$5.5 \times 10^{21}$ kg. For FeO$^{1.5}$, a maximum of $(24/56) \times 5.5 \times 10^{21} = 2.3 \times 10^{21}$ kg of O can be oxidized by $Fe^{3+}$. Given that the mass of Earth's atmosphere (1 bar) is $5.1 \times 10^{18}$ kg, nearly 500 bars of O can be taken up by this magma ocean. Although the specifics depend on the temporal evolution and other magma ocean characteristics, including size and temperature, large amounts of atmospheric oxygen can theoretically be removed in this manner, possibly explaining the lack of remnant oxygen in the Venusian atmosphere.

Moreover, the empirical HZ would shrink if Venus had lost its surface ocean by the end of accretion. As shown in Ramirez [22], this 'early Venus limit' would be computed at ~4.56 Ga, when solar luminosity was ~ 70% that of today [48]. The effective solar flux ($S_{EFF}$) at Venus' orbit at that time was 0.7x1.92 = 1.35, which corresponds to an orbital distance today of d = $\sqrt{1/0.35}$ = 0.86 AU (equation 3). Thus, our solar system's empirical HZ would shrink in size by ~0.11 AU (~10%) if Venus had lost its water early.

*3.2. The fate of Mars*

Venus lost whatever carbonate-silicate it may have had and on Earth this cycle is sustained by plate tectonics. Mars had recently exhibited hot spot volcanism, similar to Hawaii [56], but hot spot volcanism is unlikely, by itself, to regulate surface temperatures over geologic timescales. This is consistent with the absence of standing bodies of water on present Mars. The differences in bulk sizes between Earth and Mars may also be key to explaining the divergent evolutionary histories between the two planets. Mars has less volume available for its surface area, causing it to lose heat more rapidly than our planet. As time proceeds, the Martian dynamo weakens and the effects of solar wind-driven hydrodynamic escape intensify atmospheric escape processes [57]. This is consistent with an early shutdown of the Martian dynamo (~4.1 Ga; [58][59]). Although other work suggests that the dynamo may have lasted into the early Hesperian (~3.7 Ga; [60]), these authors had linked the oldest visible surface age of volcanism to the magnetization age of the crust, which may be a less accurate technique [59].

It is thought that Mars never had plate tectonics, and even if it did, it did not last very long [61,62]. Whereas Mars is likely too small (mass ~10% that of Earth's) to have supported plate tectonics over billion year timescales, plate tectonics may be favored for planets between 1 and 5 Earth masses [63]. However, whether normal or shear stresses govern the initiation of plate tectonics may complicate the above picture [64]. At even larger masses, increased resistive forces under high gravity may reduce subduction tendency (e.g., [63,65]), although this remains highly debated (e.g., [66]).

The lack of plate tectonics and (possibly) a carbonate-silicate cycle on early Mars would seem to contradict the widespread evidence of surface fluvial features suggestive of a once long-lived warmer and wetter climate (e.g., [67][68]) that was possibly capable of supporting surface oceans (e.g., [69–71]). However, it is possible that other volatile recycling mechanisms operate on other planets. For instance, $CO_2$ on early Mars could have been cycled through vigorous volcanic outgassing and subsequent melting and remobilization of surface carbonates [72]. Plate tectonics may not have even been present on the early Earth ~3.5 Ga [73] and yet our planet may have been habitable by ~4.4 Ga [74]. During the Archean, Earth may have had a form of tectonics dominated by plumes [73]. Also, although the study was performed for Earth-sized planets, if radiogenic production or volcanism is high enough, habitable conditions lasting billions of years may be sustained on stagnant lid exoplanets [75]. Thus, the possible lack of plate tectonics on early Mars may not be inconsistent with surface conditions capable of supporting standing bodies of liquid water.

**4. Limit cycles**

Limit cycles are another idea related to the cycling of $CO_2$ on planets [76,77]. Assuming that $CO_2$ is efficiently transferred between the surface and mantle through a global $CO_2$ cycling mechanism, warm planetary periods occur when volcanically-outgassed $CO_2$ outpaces drawdown, silicate weathering, and incorporation into rocks. Once the opposite occurs, and these removal processes outpace volcanic outgassing, the planet freezes instead. The process of cycling into and out of such warm conditions is called a "limit cycle" (Figure 4). Planets that receive low levels of stellar insolation, like those near the outer edge of the habitable zone, and with low enough $CO_2$ outgassing rates, are susceptible to these unstable climates [78,79]. Limit cycle frequency is a strong function of the soil $CO_2$ pressure and the volcanic outgassing flux [79]. Recent 3-D calculations [80] also confirm that limit cycles could occur on outer edge exoplanets as initially predicted in 1-D studies [79].

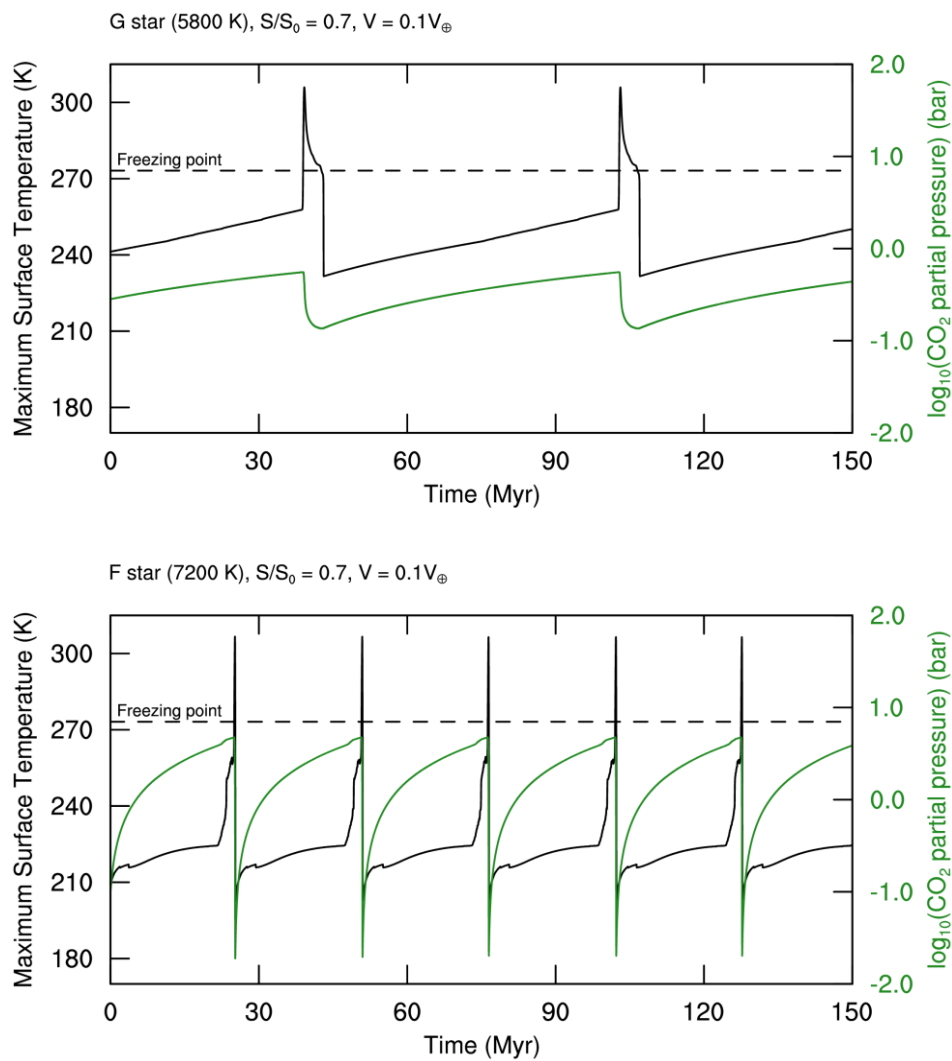

**Figure 4:** Maximum surface temperature (black curve) and $pCO_2$ level (green curve) for an Earth-like planet with a volcanic outgassing rate 10% that of Earth's orbiting a G-star (top) and F-star (bottom) for $S_{EFF}$ =0.7. Transient periods of warming during which temperatures exceed the freezing point of water (dashed line) are followed by extended periods of global glaciation (adapted from Haqq-Misra et al.[79]).

Haqq-Misra et al. [79] argued that this mechanism decreases HZ width on planets with low volcanic outgassing rates. Although it may be possible that limit cycles occur on some exoplanets [76,80], this particular result is a direct prediction of the carbonate-silicate cycle and does not depend on the occurrence of limit cycles.

This can be demonstrated by employing a recent weathering rate parameterization [81] and assuming that volcanic outgassing and weathering rates are equal at steady state (equation 5):

$$\frac{W}{W_{EARTH}} = \left(\frac{pCO_2}{p_{EARTH}}\right)^\beta e^{[k_{act}(T_{surf}-288)]} \left[1 + k_{run}(T_{surf}-288)\right]^{0.65}, \quad (5)$$

Here, $W$ is the weathering rate, $k_{act}$ is an activation energy (0.09), $T_{surf}$ is surface temperature, $k_{run}$ is a runoff efficiency factor (0.045), and $\beta$ is the dependence of $pCO_2$ on W. $W_{earth}$ and $P_{earth}$ are the soil weathering rates and soil $pCO_2$ values, respectively, for the Earth. As predicted for planets that have an operational carbonate-silicate cycle, this parameterization assumes that weathering and volcanic outgassing rates scale with pressure. I assume soil $pCO_2$ is 30 times that of the atmosphere, following ref. [78]. I also use a β of 0.35, which is consistent with experimental measurements for silicate rocks [82–84]. According to Figure 2, the $CO_2$ partial pressure at the outer edge is 8 bar and $S_{EFF}$ is 0.36 (d = $\sqrt{1/0.36}$ = 1.67 AU; from equation 3), which requires a volcanic outgassing rate ~4 times that of Earth's to maintain a mean surface temperature of 273 K (according to equation 5). At a $S_{EFF}$ of 0.6, d is ($\sqrt{1/0.6}$) ~1.3 AU, which corresponds to an atmospheric $CO_2$ partial pressure of ~0.08 bar (Figure 2). To support the same mean surface temperature of 273 K, the volcanic outgassing rate would only need to be ~1/5 as high, and lower than the Earth's (Earth's higher mean surface temperature requires larger volcanic outgassing rates). Thus, irrespective of the existence of limit cycles, planets with low volcanic outgassing rates can only support low atmospheric $CO_2$ pressures, requiring the planets to be closer to their stars to maintain warm surface conditions. Likewise, planets with high volcanic outgassing rates are located farther away at the same mean surface temperature, which is consistent with carbonate-silicate cycle predictions [33].

Although the possibility of limit cycles on some outer edge exoplanets remains intriguing, and would require observations to confirm, limit cycles likely did not occur on our planet because volcanic outgassing rates and solar insolation are both thought to have been high enough to support warm stable climates [76]. It is important to note that limit cycles should not be equated with snowball Earth episodes, which are temporary excursions into a globally frozen state thought to have occurred a few times throughout Earth's history [85,86]. Whereas limit cycles are caused by energy fluctuations in the cycling of $CO_2$, other external factors, including life, may have been the trigger for snowball Earth events. For instance, methanogenic production of $CH_4$ may have helped combat the faint young Sun problem on the early Earth until oxygen levels rose, which would have caused anaerobic methanogens to perish or become confined to restricted habitats [87].

These limit cycles have also been proposed as a possible transient warming mechanism for early Mars [78]. However, recent studies argue that limit cycles are unlikely to have occurred on early Mars for several reasons. First, observations of ancient terrains reveal no evidence of widespread glaciation [68,88,89], which is crucial for this mechanism. It is also difficult to transiently warm a very icy early Mars because melting the highly-reflective ice would require atmospheric $CO_2$ pressures that exceed available paleopressure constraints (e.g., [90–92]). At high enough ice cover, the atmosphere can collapse [68,93]. Moreover, Batalha et al. [78] had assumed a linear dependence between silicate weathering and the dissolution of H+ in ground water, which made their model especially susceptible to limit cycles. However, silicate rocks exhibit a fractional order dependence [82–84], which greatly reduces the tendency for limit cycles, while favoring warm stable solutions instead [68,93]. Finally, the Martian southern highlands, where the valley networks are located, are several kilometers higher than the global topographic mean, making them particularly prone to glaciation if the early Martian climate was cold and icy (e.g., [39,46]). However, the models that predict limit cycles for early Mars have assumed a flat topography and lack a hydrologic cycle [78]. In comparison, more detailed models find that $CO_2$ is likely to condense en masse at the poles and in the Martian highland regions on a cold and icy early Mars, causing the ice to become thick and stable against deglaciation [94].

Radiogenic heat production (and volcanism) on planets is expected to decrease over geologic timescales as star formation decreases [14]. For instance, if an Earth twin had formed today its radiogenic heat production would be ~60% lower in 4.5 billion years than the heat budget of our present Earth [14]. This may increase limit cycle tendency for such planets although the effect of the brightening Sun would offset this tendency.

## 5. The role of oxygen in habitability

The search for water is accompanied by the search for oxygen in planetary atmospheres. This section reviews the latest advances in our understanding of this key gas. I also discuss its relation to the HZ and make some assessments.

*5.1 The importance of oxygen-poor planets in the search for extraterrestrial life*

One reason why oxygen is thought to be essential to complex life here and elsewhere is that all known large animals require it. Moreover, the Cambrian explosion at ~ 0.54 Ga, an event associated with the rise of large complex lifeforms, follows a purported rise in oxygen levels at ~0.6 Ga (e.g., [95]).

Oxidative metabolisms may also be the most efficient ones for complex life to use because resultant energy yields are much higher than for anoxic equivalents(e.g., [96]). With the recent discovery of Locifera, however, it now seems possible for complex life, at least relatively small organisms, to live their entire lives in the complete absence of oxygen [97]. This is because Locifera are small metazoans that do not need the high energy yields apparently required by larger lifeforms, which allows them to survive with simpler anoxic metabolisms.

It is still uncertain whether the rise of oxygen at ~0.6 Ga directly led to the Cambrian Explosion, or if other factors were necessary. For instance, the Great Oxidation Event (GOE) between ~1.6 – 2.5 Ga (2.33 Ga current best estimate [98]) generated relatively high oxygen levels (on at least the percent level) and yet life remained relatively small [99]. Indeed, oxygen levels during the GOE may have briefly rivaled those today even though complex life was absent [100].This would indicate that evolution towards large animals may not require long (~4 billion year) oxygenation times [96]. Indeed, some large animals (sponges) can survive on much lower oxygen levels (< 1% PAL) than previously thought [99]. Plus, cnidaria and porifera (sponges) already existed hundreds of millions of years before the Cambrian Explosion and at different times [101,102]. Molecular data also support the notion that metazoan diversification occurred well before and into the Cambrian explosion [103,104]. All of this suggests that the increase in animal complexity was not a sudden event, but a series of incidents that occurred over a significant amount of geologic time (e.g., hundreds of millions of years) [105]. Instead of a rise of oxygen, perhaps genetic diversification during the Vendian allowed for the gradual development of complex animal forms that triggered the Cambrian Explosion [105]. Such scenarios would indicate that oxygen, at best, may only be a facilitating condition, but not the defining one for the evolution of animal life [105].

Nevertheless, in many astrobiological discussions, the focus in the search for atmospheric biosignatures remains largely a search for biotic oxygen (and $CH_4$), which includes finding both simple and complex life (and possibly intelligence) elsewhere (e.g., [106,107]). However, an extraterrestrial astronomer observing the Earth over time would find that oxygen levels were well under 1% for most of its history [108], possibly concluding (incorrectly) that the Earth is lifeless. $CH_4$, another molecular compound associated with life, was produced by anaerobic methanogens during much of this oxygen-poor period (e.g., [99,109]). Although alternative viewpoints exist [105], a popular opinion has been that simpler organisms may be much more common in the cosmos than is complex animal life [110]. These are all important points that should be noted, and a thorough, rather than limited approach, should be undertaken in the search for extraterrestrial life, as is argued throughout this review.

*5.2 Is oxygen really a good biosignature?*

Biotic oxygen production rates on Earth far exceed abiotic ones, suggesting that high atmospheric oxygen concentrations, perhaps in conjunction with $CH_4$, may be good bioindicators (e.g., [106,111,112]). Thus, for this hypothesis to hold, abiotic mechanisms should either be unable to produce oxygen efficiently or (at the least) be easily distinguished from biotic oxygen production processes.

One way to potentially produce significant amounts of abiotic oxygen is via early hydrodynamic escape, as could have occurred on Venus (although a magma ocean may have removed excess atmospheric oxygen) (e.g., [113] and see Section 3.1). Ignoring surface sinks, Luger and Barnes [114] applied this idea to planets in the main-sequence habitable zones of M-dwarfs, finding that hundreds – thousands of bars of abiotic $O_2$ could be produced during accretion. However, allowing H and O inventories to evolve over time, in contrast to Luger and Barnes [114], Tian [115] found that planets with Earth-like water inventories are likely to exhibit almost no oxygen buildup, even in the absence of surface sinks. This result has been subsequently confirmed (e.g., [55,116]). Abiotic oxygen buildup appears possible on Earth-sized planets that contain at least several percent of their water by mass [55]. For sub-Earths (~ Mars-sized), oxygen buildup is also extremely unlikely [117]. Thus, significant abiotic oxygen buildup with this mechanism seems implausible under many circumstances (ibid).

Another way to produce high levels of abiotic oxygen is via $CO_2$ photolysis, which liberates O atoms that recombine to form $O_2$ and $O_3$ (e.g., [118–122]). However, the resultant concentrations are well under 1%, which would argue against an unambiguous biotic signature (although we are in danger of concluding that there is no life when there may well be. See Section 5.1). Alternatively, recombination reactions that form $CO_2$ may not proceed on planets with little water, leading to considerable $O_2$ buildup [123]. However, such dry worlds are easily distinguished by the lack of a strong water vapor signal.

Yet another way to potentially produce abiotic oxygen is via enhanced water vapor photolysis in the upper atmosphere (i.e., the cold trap), leading to the preferential escape of lighter H molecules to space [24]. This concept is essentially analogous to the "moist greenhouse" of Kasting [37] except that the moist greenhouse is the specific case in which enhanced water vapor photolysis can desiccate a planet with an Earth-like water inventory within the age of the solar system (as discussed earlier). On planets with lower non-condensible inventories than the Earth, this enhanced photolysis occurs at relatively lower temperatures (< 300 K) whereas it is triggered at higher surface temperature (> 350 K) for planets with larger non-condensible inventories (e.g., [1,24]).

Thus, it has been argued that planets undergoing such enhanced water vapor photolysis in their upper atmospheres are uninhabited because of high abiotic oxygen levels (e.g., [106]). Although, the mean surface temperatures for an Earth clone undergoing a moist greenhouse (~340 K) are detrimental for complex animal life, they are still well under the hottest temperatures that can be supported for terrestrial microbial life [11]. Furthermore, mean surface temperatures for moist greenhouse planets with low non-condensible inventories are low enough to comfortably support animal life. Therefore, it isn't clear that significant abiotic oxygen buildup levels on planets with low non-condensible inventories require that life be absent. More observations would be needed to make conclusive determinations should such a planet be found.

Overall, current research suggests that high levels of atmospheric oxygen may be a robust biosignature, especially once hydrogen species are characterized and false positives are ruled out (e.g., [106]). Nevertheless, there are caveats with inferring potential life from observed high atmospheric oxygen levels [112]. At high enough oxygen partial pressures, breathing oxygen strips molecules of electrons, creating free radicals that destroy cells, causing DNA damage in humans and other lifeforms on Earth (e.g., [124]). Although such cell damage is slow-acting in Earth's lifeforms at the current oxygen concentration, they could be debilitating, even fatal, at higher pressures (e.g., [125]). Plus, wildfires on vegetated planets may limit $O_2$ abundance and act as a negative feedback when $O_2$ levels rise too high [126,127]. Nevertheless, aerotolerant anaerobes on Earth employ fermentation to produce Adenosine triphosphate (ATP), which is a complex organic chemical, and are not negatively affected by the presence nor absence of oxygen (e.g., [128]). Thus, the maximum oxygen pressure level sustainable by alien life is unknown.

## 6. Expanding the spectral range of the classical HZ

Given the various uncertainties in our current understanding (as explained above), recent work has instead adopted a more liberal view of the HZ (e.g., [2,51,129–132]), expanding it to include the detection of life beyond that suggested by the classical definition. As explained in the next section, that would also include atmospheres consisting of reduced gases like $CH_4$ or $H_2$, ocean worlds, desert worlds, and even planets orbiting stars that are still in the pre-main-sequence (or post-main-sequence) phase of stellar evolution, and many more possibilities that will be discussed later.

Previous calculations of the classical HZ had only computed boundaries for stars with main-sequence lifetimes > ~1 or 2 Ga (F- M spectral classes)[1,26]. However, some argue (e.g., [2,130]) that such a HZ (e.g. [1]) may still be overly-pessimistic. For instance, we know that microbial life on our planet arose much more quickly than that (~ 700 Myr; [133]). Thus, main-sequence stars that last at least 700 Myr (which includes A-stars with $T_{EFF}$ up to ~10,000 K) should be included in HZ definitions (e.g., [2,130,131]). After all, if a planet can sustain simple life it is habitable (by definition), as originally defined by Kasting [1]. Again, it is possible that life emerges even more quickly than 700 Myr on other planets. For instance, evidence in zircons suggests that conditions on Earth may have been habitable by 4.3 – 4.4 Ga [74]. Overall, our own planet exhibits evidence that life can arise *at least* as quickly as 700 Myr, if not quicker.

Classical HZ boundaries were originally computed for stars of $T_{EFF}$ = 2,600 – 7,200 (F – M stars) using the following fourth-order parameterization of the stellar effective flux ($S_{EFF}$)[26]:

$$S_{EFF} = S_{sun} + a \cdot T^{*} + b \cdot T^{*2} + c \cdot T^{*3} + d \cdot T^{*4}, \qquad (6)$$

Where $T^* = (T_{EFF} - 5780)$ and $S_{sun}$ is the $S_{EFF}$ value for a given HZ limit in our solar system. The quantities (a,b,c,d) are constants. However, the spectral range has recently been expanded to include A-stars (10,000 K) [2]. This extended classical $CO_2$-$H_2O$ HZ, along with updated constants for equation 6 (Table 1), is shown in Figure 5.

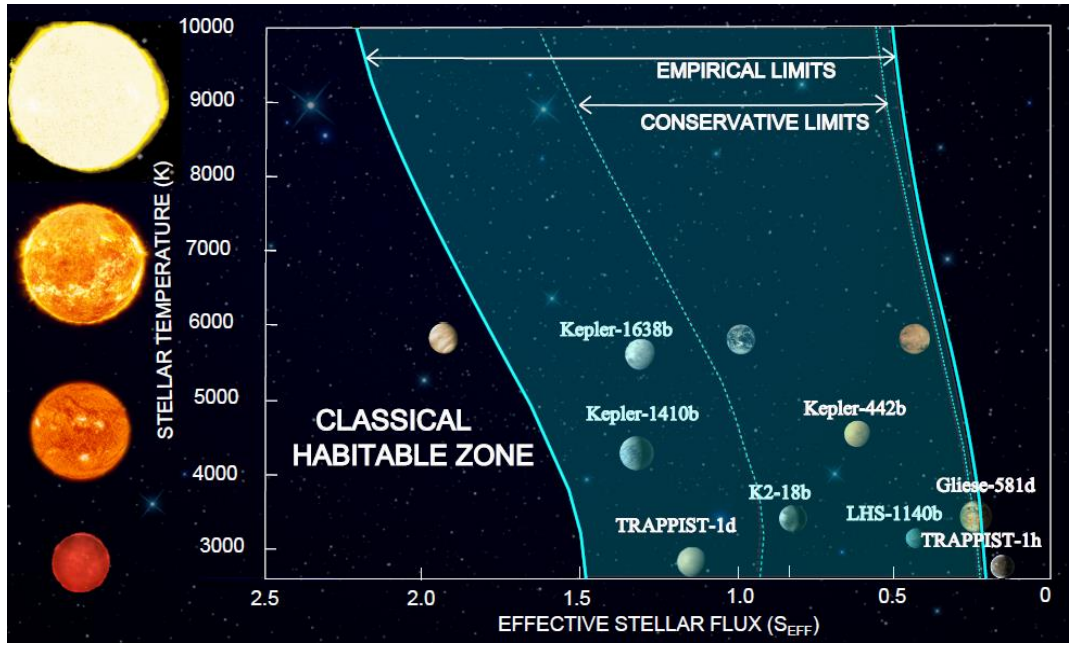

**Figure 5:** Extended classical habitable zone for stellar temperatures from 2,600 to 10,000 K (based on work from Ramirez and Kaltenegger [2])

**Table 1.** Revised coefficients for the extended classical HZ (atmospheric composition $N_2$-$CO_2$-$H_2O$) and the empirical HZ for host stars with $T_{EFF}$ = 2,600 – 10,000 K*

| Constant | Recent Venus | Leconte et al. | $CO_2$ Maximum Greenhouse | Early Mars |
|---|---|---|---|---|
| $S_{EFF(Sun)}$ | 1.768 | 1.1105 | 0.3587 | 0.3246 |
| A | $1.3151 \times 10^{-4}$ | $1.1921 \times 10^{-4}$ | $5.8087 \times 10^{-5}$ | $5.213 \times 10^{-5}$ |
| B | $5.8695 \times 10^{-10}$ | $9.5932 \times 10^{-9}$ | $1.5393 \times 10^{-9}$ | $4.5245 \times 10^{-10}$ |
| C | $-2.8895 \times 10^{-12}$ | $-2.6189 \times 10^{-12}$ | $-8.3547 \times 10^{-13}$ | $-1.0223 \times 10^{-12}$ |
| D | $3.2174 \times 10^{-16}$ | $1.3710 \times 10^{-16}$ | $1.0319 \times 10^{-16}$ | $9.6376 \times 10^{-17}$ |

*Reproduced from Ramirez and Kaltenegger [2] and fixing a typo.

**7. Planetary habitability: extensions in space**

In addition to an incomplete understanding of terrestrial or extraterrestrial biology, our technological capabilities for the search are limited [134]. One estimate suggests that upcoming direct imaging missions using an 8-m telescope would only find ~ 20 – 30 Earth-sized planets out of some 500 stars (~4 – 6%) in the main-sequence HZ, although this number can be ~ a factor of 2 higher with more optimistic assumptions [135]. It is unclear whether such exoplanet yields are sufficient to result in a successful search and so it would be prudent to think as broadly about the possibilities for life as we can. The following sections review recent advances in HZ theory, which can be used to improve our chances of finding extraterrestrial life.

*7.1. Extending the habitable zone with hydrogen*

The HZ is an atmospheric composition-dependent concept whose boundaries greatly depend on the exact mix of greenhouse gases considered (e.g., [3]). The classical $CO_2$-$H_2O$ HZ in our solar system is located where it is ( ~ 0.75 – 1.77 AU (empirical) or ~0.95 – 1.67 AU (conservative)) because the greenhouse effect becomes inefficient at the $CO_2$-dominated outer edge whereas water loss to space imperils habitability at the $H_2O$-dominated inner edge. However, real planets (like Earth) contain a slew of additional greenhouse gases, including $CH_4$, $H_2S$, $SO_2$, $H_2$ and many others. The consideration of such different atmospheric absorbers will thusly lead to boundaries that can differ greatly from those predicted by the classical definition.

The extreme case would be to forego the $CO_2$-$H_2O$ concept entirely and compute an alternate HZ using completely different greenhouse gas combinations. Pierrehumbert and Gaidos [129] suggested that young protoplanets can accrete prodigious amounts of primordial hydrogen from the protoplanetary disk, similar to previous suggestions for orphan planets and early Mars [136,137]. Pierrehumbert and Gaidos [129] predicted that a super-Earth with a 40 bar primordial hydrogen atmosphere could achieve warm mean surface temperatures (> 273 K) out to 10 AU around G-type stars. This potent greenhouse effect arises from collision-induced absorption (CIA) of self-broadened $H_2$-$H_2$ pairs.

However, hydrogen is a light gas, and maintaining large amounts of it over geological timescales is difficult, particularly for planets that are located at shorter distances, like those within the classical HZ. Without a continuous hydrogen source, hydrodynamic escape to space could strip the dense primordial hydrogen atmospheres of rocky classical HZ planets in just a few to ~ 100 million years [138]. Alternatively, perhaps habitability on such worlds can be sustained if biological feedbacks that can regulate $H_2$ arise [139]

Another way to generate atmospheric hydrogen is through volcanism. Paleoclimate studies of early Earth and Mars show that volcanism could have outpaced $H_2$ escape on both planets, maintaining clement conditions with $H_2$ partial pressures as low as < 1 bar [46,93,140,141]. This is because reduced mantle conditions could have favored enhanced outgassing of $H_2$ over longer timescales. Unlike the previously described primordial hydrogen mechanism [129], hydrogen is not the major constituent, as losses to space are compensated by volcanism. In this mechanism, the background atmosphere *foreign-broadens* the hydrogen which excites roto-translational bands, absorbing in additional spectral regions where $CO_2$ and $H_2O$ absorb poorly [93,140,142–144]. On Earth, $N_2$ was suggested to be this background gas [46] whereas for Mars it could have been $CO_2$ and $N_2$ [93,140]. Nevertheless, $CO_2$-$H_2$ CIA is even stronger than that for $N_2$-$H_2$ CIA, requiring lower pressures to achieve warm surface conditions for the former atmospheric composition than the latter [141].

Although Earth's mantle is thought to have oxidized early, making it unlikely that this mechanism had occurred on our planet [145], empirical data from Martian meteorites (including NWA Black Beauty and ALH84100) suggest a highly-reduced early martian mantle that promoted voluminous hydrogen outgassing (e.g., [146]), possibly for over 0.5 billion years [132,140]. The reason why Earth's mantle may have oxidized early (during accretion) is because pressures in the lower mantle were sufficiently high to convert iron(II) oxide to iron metal and iron(III) oxide [147]. This would argue against the $H_2$-rich early Earth of Wordsworth and Pierrehumbert [148]. However, small planets like Mars would not have generated sufficiently high internal pressures to trigger such mantle oxidation, which is consistent with the meteoritic evidence.

On early Mars, high $CO_2$ concentrations under a highly-reducing mantle would have been sustained as $H_2O$ vapor reacted with outgassed reduced gases, like CO and $CH_4$, in addition to what may have survived primordially (e.g., [22,140]). According to recent calculations [93,141], ~ 3% $H_2$ and less than 2 bars of $CO_2$ could have sustained warm surface conditions on early Mars.

Ramirez and Kaltenegger [132] had extended the classical HZ width by borrowing their $CO_2$-$H_2$ CIA idea for early Mars (i.e., Ramirez et al. [140]) to compute a $N_2$-$CO_2$-$H_2O$-$H_2$ HZ. They found that the outer edge for our solar system would extend from ~ 1.67 – 2.4 AU, assuming a hydrogen concentration of 50%. Similar extensions to the outer edge were found for A – M spectral classes. In contrast, the inner edge was only minimally affected (~0.1 – 4%) because $H_2$ warming becomes small in $H_2O$-dominated atmospheres. Unlike warmth via dense primordial hydrogen atmospheres [129], this mechanism provides a continuous hydrogen source. So long as volcanic outgassing can outpace escape, habitable conditions may be sustained on geologic (e.g., > 1 billion year) timescales. Another advantage of the volcanic hydrogen HZ over the classic definition is that $H_2$, being a light gas, increases the atmospheric scale height, facilitating the detection of bioindicators in transit spectroscopy. For example, adding 30% $H_2$ to a habitable planet's atmosphere located at ~1.7 AU increases its atmospheric scale height by over 60% (assuming a similar temperature structure) [132]. Some planets that are outside the classical HZ, like TRAPPIST-1h, are located inside the volcanic hydrogen HZ and may be potentially habitable [149].

Although this mechanism may work best for small terrestrial planets, some large terrestrial worlds could potentially support high atmospheric concentrations through their higher gravity and potentially stronger magnetic fields (e.g., [129,132]). Also, early hydrodynamic escape rates are likely to be lower on planets near the outer edge of the HZ (e.g., [50]).

A criticism with both the primordial hydrogen (e.g., [129]) and volcanic hydrogen (e.g., [132]) greenhouse mechanisms is the doubt that such planets could support life. On Earth, free hydrogen would be readily consumed by methanogens, so high hydrogen concentrations may suggest an absence of life (e.g., [150]). However, such arguments are based on how life evolved on our own planet, so it is almost impossible to predict how life would evolve on worlds that have a completely different redox budget. Alternatively, extraterrestrial organisms could have evolved a hydrogen-based form of photosynthesis, one that also evolves by consuming hydrogen and producing it back out [19]. In this case, the biotic hydrogen produced may be hard to distinguish from that of the background gas (ibid). However, seasonal or spatial variations in the $H_2$ concentration is one way in which life may be potentially inferred in this case, particularly if productivity is associated with temperature as per Daisyworld [2]. Similar arguments had been used to suggest that methane on present Mars may have a biotic origin [151].Ammonia is another potential biosignature gas in such atmospheres [19,152], although (like oxygen) detailed observations would be required to eliminate potential false positives [153] . In these anoxic atmospheres, life may be small and would not need the high energy requirements of oxygen-based life. I will return to these points in Section 15.

*7.2. Extending the habitable zone with methane*

Previous HZ studies have shown that the stellar energy distribution (SED) of a star influences both the magnitude and location of atmospheric warming (e.g., [1,26,51,131,154–156]). In comparison to HZ planets orbiting G-stars, M-star HZ planets absorb more energy whereas those around hotter (F and A) stars absorb less. This is because 1) Rayleigh scattering decreases at longer wavelengths, and 2) near-infrared absorption in planetary atmospheres is higher at the longer peak wavelength energies emitted by M-stars (Figure 6)([1]). Thus, the same integrated stellar flux that hits the top of a cool star's planetary atmosphere is warmed more efficiently than the same flux from a hot blue star.

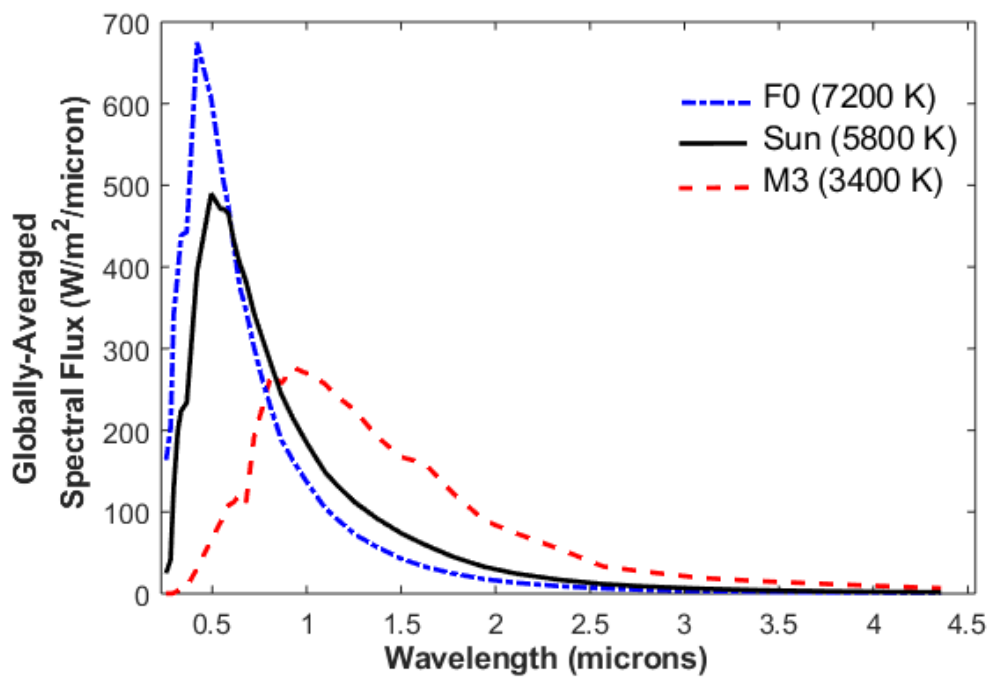

**Figure 6**: Globally-averaged incoming stellar spectra for the Sun, F0, and M3 stars. Adapted from Ramirez and Kaltenegger [2] and updated from Ramirez [22].

However, Ramirez and Kaltenegger [2] and Ramirez [22] showed that the addition of $CH_4$ to classical $CO_2$-$H_2O$ HZ atmosphere produces net greenhouse warming (tens of degrees) in planets orbiting stars hotter than a mid-K (~4500K), whereas this causes a prominent anti-greenhouse effect in planets orbiting cooler stars (Figure 7). Although a 1% $CH_4$ concentration is a reasonable upper bound estimate for the early Earth (e.g., [157]), there is no reason to believe that higher concentrations are not possible on other planets. Assuming a maximum $CH_4$ concentration that is 10% that of $CO_2$, above which organic $CH_4$ hazes form that can cool the climate (e.g., [158]), the outer edge distance can increase by over 20% for the hottest stars ($T_{EFF}$ = 10,000 K) whereas it *decreases* by up to a similar percentage for the coolest stars ($T_{EFF}$ = 2600 K)[2]. For our solar system, the outer edge distance increases from 1.67 to 1.81 AU (~8% increase)[2].

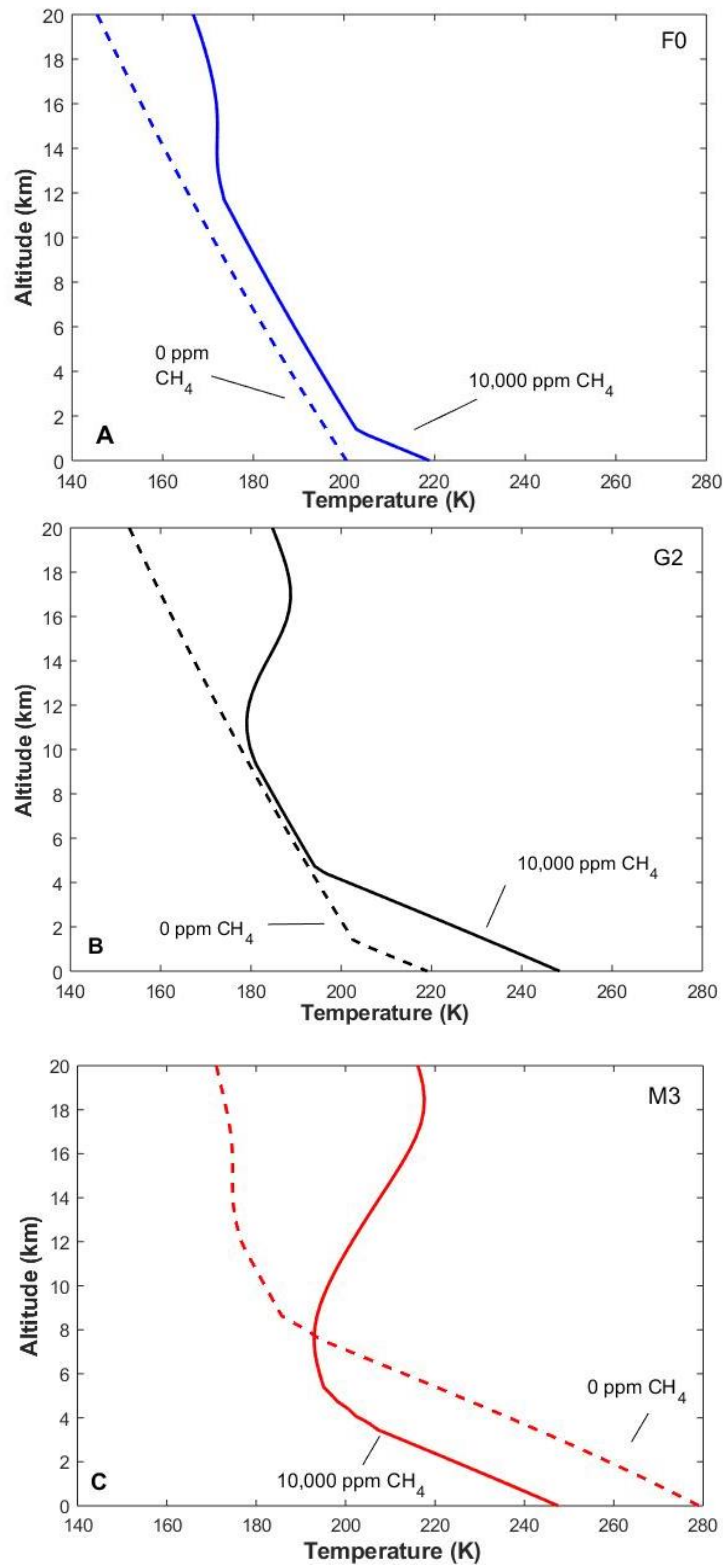

**Figure 7:** Changes in temperature profiles for a 3 bar $CO_2$ (1 bar $N_2$) atmosphere (solar energy =33% that received by the Earth) for a model planet orbiting an F0 (blue), b) a G2 (black; solar-analog), and c) an M3 (red) host star with and without 1% (10,000 ppm) $CH_4$. A temperature inversion forms for all three planets. The surface temperature increases when $CH_4$ is added on the planet orbiting the F0 (18K) and solar (G2) analog (29K), but decreases for the planet orbiting an M3 star (31K). Adapted from Ramirez and Kaltenegger [2] and updated from Ramirez [22].

Methane cools planetary atmospheres orbiting cooler stars because their higher near-infrared emission levels preferentially heat the upper atmosphere, producing large stratospheric temperature inversions that counteract the thermal infrared greenhouse effect (Figure 7) [2,22,159]. A recent study argues that haze-induced cooling on M-star planets may be insignificant [160]. However, the cooling from $CH_4$ absorption would still occur because this mechanism is independent of any haze formation [2]. A similar cooling behavior from $CH_4$ was also recently confirmed in 3-D simulations of the TRAPPIST-1 planets [161]. In contrast, at stellar effective temperatures above ~4500 K, the greenhouse effect outstrips upper atmospheric heating, producing net warming [2]. The extended $CO_2$-$CH_4$-$H_2$ HZ for stars of stellar effective temperatures between 2,600 and 10,000 K is given in Figure 8.

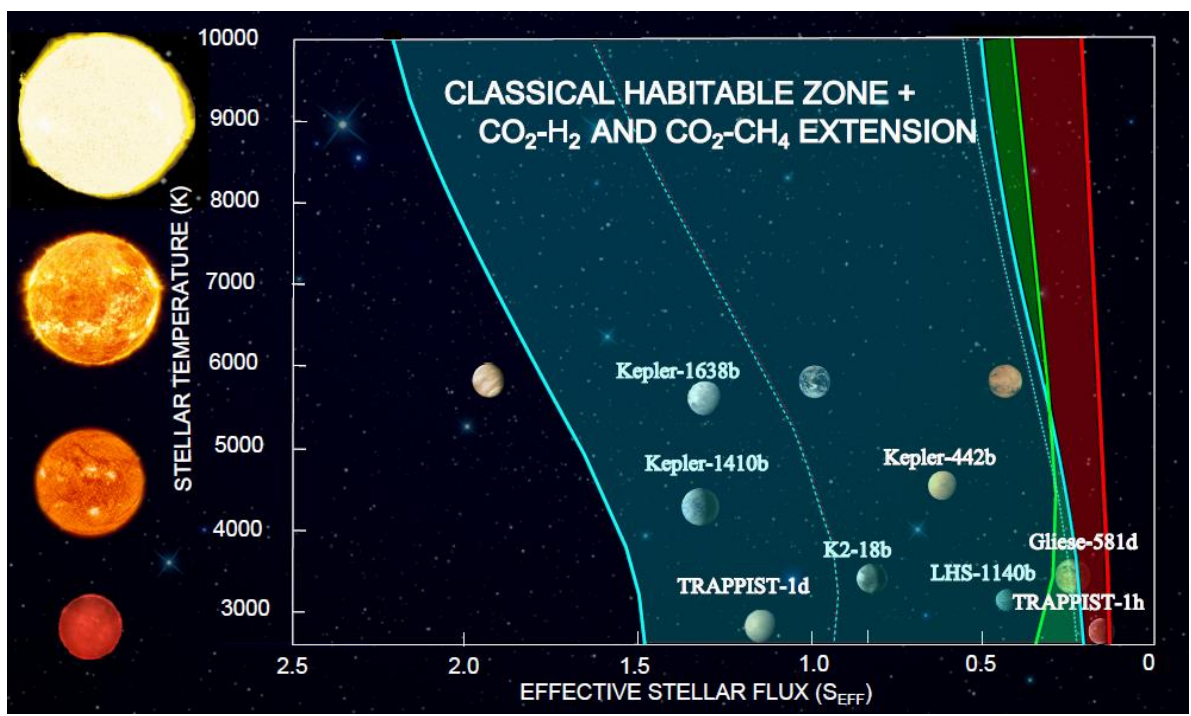

**Figure 8**: The classical HZ (blue) with $CO_2$-$CH_4$ (green) and $CO_2$-$H_2$ (red) extensions for stars of stellar temperatures between 2,600 and 10,000 K (A – M-stars). The same classical HZ boundaries shown in Figures 3 and 5 are also shown. Some solar system planets and exoplanets are also shown. Based on work from Ramirez and Kaltenegger [2,131,132] and Ramirez [22].

Atmospheric methane can be abiotically produced in several ways, including via volcanism, serpentinization, impact processes, and hydrothermal activity (e.g., [2,162–164]). High $CH_4$ concentrations via volcanism are unlikely unless the mantle is highly-reduced, as per $H_2$ (discussed above). Impacts can erode or enhance planetary atmospheres, depending on impactor and target material properties [162]. Methane can also be produced by serpentinization, which is a process by which Fe-rich waters produce $H_2$ via oxidation of basaltic crust (e.g., [165]). On Earth, the rate of methane production via this process can be bounded by the amount of seafloor that can be oxidized [166]:

$$CO_2 + 2H_2O \leftrightarrow CH_4 + 2O_2, \qquad (7)$$

Serpentinization of seafloor on Earth produces ~ $2\times10^{11}$ moles of $O_2$ per year [166]. Assuming that serpentinization only produces $CH_4$ (not $H_2$), an upper bound of $1\times10^{11}$ moles of $CH_4$ or $4\times10^8$ molecules/cm²/sec are produced on the Earth according to the above equation. However, if rocks are ultra-mafic, serpentinization rates can be quite high in localized regions, ~$1\times10^{12}$ – $1\times10^{13}$ molecules/cm²/sec for early Mars conditions [167]. To produce the aforementioned $CH_4$ production rates, serpentinization would need to occur on ~0.13 – 1.3% of a planet's surface area [2].

Although abiotic sources of $CH_4$ sustaining these dense $CO_2$-$CH_4$ atmospheres are rather significant, major sinks include photolysis and atmospheric escape. Assuming diffusion-limited escape and a homopause $H_2$ mixing ratio of 0.5%, escape rates were found to be ~ $10^{10}$ – $10^{11}$ molecules/cm²/sec. For photolysis, an estimate for the maximum photodissociation rate for a planet at 1.8 AU around a G2 star was ~$3.6 – 6.4\times10^{10}$ molecules/cm²/sec [2].

Thus, $CH_4$ losses were found to be comparable to abiotic $CH_4$ production rates for dense $CO_2$-$CH_4$ atmospheres near the outer edge. It is challenging for abiotic mechanisms alone to support such high $CH_4$ concentrations. Although blue stars (A- and F-class) have decreased Lyman-alpha emission (e.g., [168,169]), and host HZ planets that are located relatively far away [2], overall photolysis rates should still be high because of enhanced emission at FUV wavelengths (e.g. [170,171]). Thus, dense $CO_2$-$CH_4$ atmospheres near the outer edge of hotter stars may suggest inhabitance and should be followed up with observations. In contrast, glaciated surfaces may characterize planets near the outer edge of colder stars [2].

*7.3. Methane Daisyworld for hotter stars*

If methane productivity increases with surface temperature, a stabilizing feedback loop suggested for the Archean Earth ala "Daisyworld" [172,173] can be proposed for habitable planets near the outer edges of hotter stars [2]. In this formulation, increases in atmospheric $CH_4/CO_2$ ratio yield a parabolic response in surface temperature whereas $CH_4/CO_2$ exhibits linear responses to increases in surface temperature (Figure 9). A positive (and therefore unstable) feedback ensues for points on the left half of the curve (e.g., $P_1$), with surface temperatures rising as $CH_4$ is increased. Such a positive feedback may operate if methanogens could evolve on these planets, assuming methane productivity increases with temperature [173]. The resultant methanogenesis yields the following, assuming atmospheric $H_2$ is available:

$$4H_2 + CO_2 \rightarrow CH_4 + 2H_2O, \qquad (8)$$

However, hazes form once $CH_4/CO_2$ ratios exceed ~0.1 [158] and the surface cools until a stable equilibrium point (e.g., $P_2$) is reached, which intersects a line with a negative (and therefore stable) slope. Further increases in surface temperature increase the $CH_4/CO_2$ ratio, thickening the haze and countering the warming. However, if more complex models confirm that high FUV fluxes inhibit haze formation on blue stars (A and F spectral class) and active M-dwarfs [160], the system could potentially achieve even higher temperatures until runaway conditions are triggered or life extinguishes itself [2].

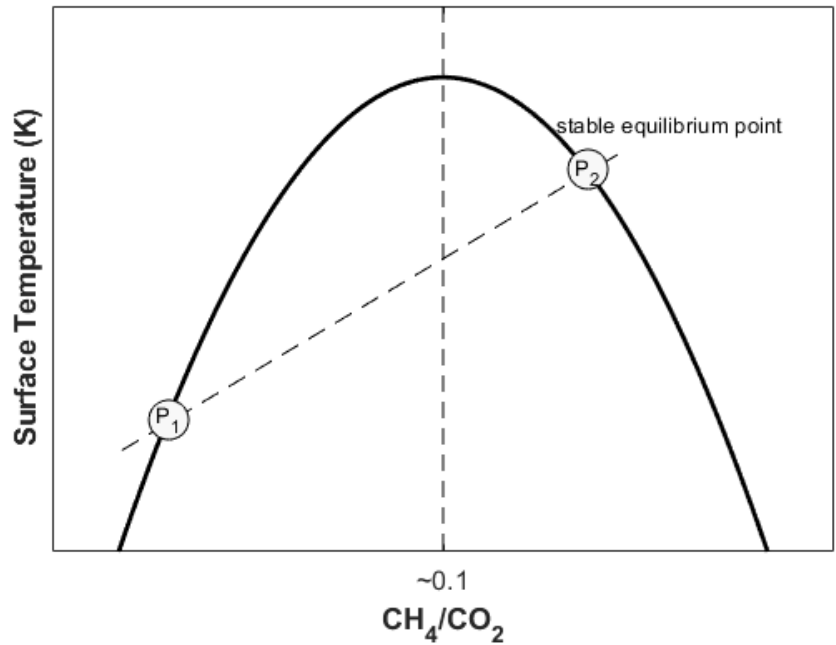

**Figure 9:** Proposed "Daisyworld" scenario for planets at the outer edge of the methane HZ with $CH_4$ in their atmospheres orbiting hotter (~A – G class) stars. The parabolic lines illustrate the effect that $CH_4/CO_2$ has on surface temperature whereas the straight lines depict the effect of temperature on the $CH_4/CO_2$ ratio (figure adapted from Ramirez and Kaltenegger [2] and Domagal-Goldman et al.[173] ).

## 8. Planetary habitability: extensions in time

Although the search for potentially habitable exoplanets has focused on worlds orbiting main-sequence stars, recent work shows that understanding the temporal evolution of the habitable zone is essential to determining a planet's habitability (e.g., [51,130,131]). Hart [36] was the first to consider the temporal evolution of the habitable zone as a star ages, which he termed the "continuous habitable zone." Subsequently, Kasting et al. [1] calculated a continuous habitable zone for stars of 0.5 – 1.5 solar masses during the main-sequence of stellar evolution. These authors had decided to ignore the temporal evolution of the post-main-sequence and pre-main-sequence habitable zones. However, as Danchi and Lopez [130] showed, the temporal evolution is important because the chances for life to be detected improves the longer that planets can stay within this extended temporal HZ. Not only is it possible for life to potentially exist during the formative and ending phases of a star's life (e.g., [51,130,131]), but an educated assessment of main-sequence habitability cannot be made without understanding these other phases [51]. Very little work has been done on these topics, but this is an exciting area with many possibilities. I will summarize recent results.

*8.1. Habitability during the pre-main-sequence*

M-dwarfs are smaller and cooler than other stars, which requires that their HZ planets be on closer-in orbits. Such tightly-packed orbits trigger strong tidal forces that gradually slow rotation rates until synchronous rotation is achieved. Although these worlds may be located within the main-sequence HZ, nightside temperatures are so cold that the major atmospheric constituent ($H_2O$ near the inner edge or $CO_2$ near the outer edge) would condense out en masse, triggering atmospheric collapse and rendering them uninhabitable [174]. However, a ~1 – 1.5 bar $CO_2$ atmosphere may be dense enough to efficiently transfer heat between the day- and nightsides and maintain habitable conditions [175,176]. Subsequent work suggests that dense enough atmospheres may also preclude synchronous rotation entirely [177]. However, all of this assumes the untested premise that life is possible in very dense $CO_2$ atmospheres (see Section 15.2). Also, such tightly-packed orbits around M-stars suggest high impactor velocities, which tends to favor the erosion of planetary atmospheres (e.g., [178,179] and see counterpoints in Section 9).

Moreover, such proximity to their host stars indicates a high radiation environment exposed to stellar winds and flares [180,181]. For instance, if Proxima Centauri b is assumed to have an Earth-like atmosphere , 1 bar of $CO_2$ can be lost in under 25 Myr, with much greater losses over geologic timescales [181]. Plus, even if the stellar radiation does not completely remove the atmosphere, the surface may be sterilized and unable to support life, much like the Martian surface today (e.g., [182] ).

Another major problem is that the pre-main-sequence stellar luminosity of M-stars is orders of magnitude higher than their main-sequence values (Figure 10). These bright young M-stars would trigger runaway greenhouses on worlds that are currently located in the main-sequence HZ, possibly desiccating them [51,114,183]. Only worlds that are in the pre-main-sequence HZ would avoid such water losses (Figure 10).

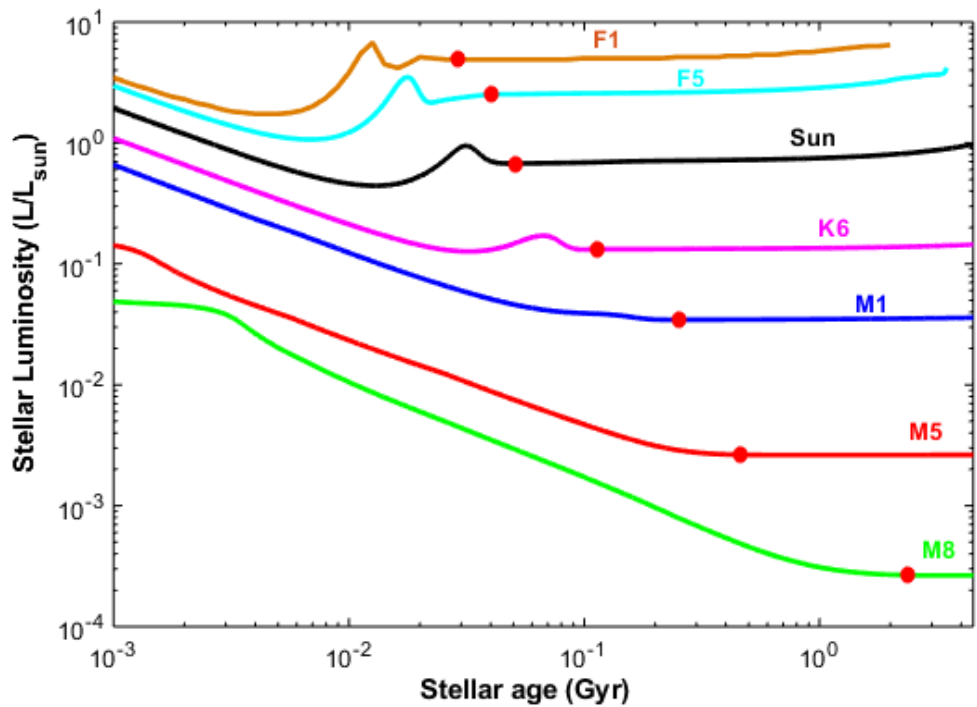

**Figure 10:** Evolution of stellar luminosity for F – M stars (F1, F5, Sun, K6, M1, M5, and M8) using Barrafe et al. [184] stellar evolutionary models. When the star reaches the main-sequence (red points) the luminosity curve flattens (updated from Ramirez and Kaltenegger [51]).

The following simplified energy limited approach illustrates the problem. In these runaway greenhouse atmospheres, water molecules in the upper atmosphere dissociate into H and O atoms [185]. Escape fluxes would be high enough for the lighter hydrogen atoms to also drag away heavier molecules (like O) along with them to space. This early phase of intense atmospheric escape would only be limited by the amount of stellar energy available to drive the flow because $H_2O$ vapor mixing ratios in these runaway greenhouse atmospheres are high enough to overcome the diffusion limit [18].

The hydrogen escape flux ($F_1(t)$)(in moles/sec) obeys the following relationship,

$$F_1(t) = 4\pi R^2 \frac{\phi(t)R}{GMm_1}\varepsilon\left(\frac{1}{d(AU)}\right)^2, \qquad (9)$$

Where $m_1$ is the mass of one hydrogen atom (1 amu or g/mol), $R$ is the planetary radius, $G$ is the gravitational constant, $M$ is the planetary mass, $\varepsilon$ is the heating efficiency (typically, 0.15), and $\phi(t)$ is the temporal evolution of the EUV flux, which is $= 29.7t[Gyr]^{-1.23}$ [186]. The above equation is then integrated to calculate moles of atmosphere lost ($Mass_{tot}$) over some elapsed time interval ($t_o$ to $t_f$) where $t_o$ is assumed to be 1 million years after the star forms and $t_f$ is the end of accretion time.

$$Mass_{tot} = \int F_1(t)dt = 4\pi R^2 \frac{\varepsilon R}{9GMm_1} B\left(\frac{1}{d(AU)}\right)^2 \int_{t_o}^{t_f} \phi(t)dt, \qquad (10)$$

However, not all the mass lost would be $H_2$. The above expression is multiplied by the $H_2$ to $H_2O$ mass ratio (1/9), given that some of the stellar energy will also be driving O escape (e.g., [51]). Equation 10 is also multiplied by the conversion factor for Gyr to seconds (B). Alternatively, the above equation could be rewritten with a reference escape flux that sums the mass flux contributions of H and O atoms (e.g., [113,114,187]). Moreover, infrared coolants, which are not present in any of these simpler models (ibid), would also lower escape rates. Despite the above caveats, this equation is sufficient for the illustrative purposes shown below.

If we assume that Venus was in a runaway greenhouse state between t =1 and 7 Myr and t = 20 – 50 Myr, according to one model prediction (Figure 4 in Ramirez and Kaltenegger [51]), equation 10 predicts that ~$3x10^{23}$ moles of $H_2$ (1 Earth ocean contains $7.8x10^{22}$ moles of $H_2$) or ~4 Earth oceans could have been lost. In contrast, if Earth was in a runaway greenhouse state, this may have occurred only briefly in the very beginning (t = 1 – 2Myr). Although Earth may have lost ~ 1 Earth ocean during this time, the Ramirez and Kaltenegger [51] model predicts that Earth was not in a runaway greenhouse state during the final stages of accretion when most of the water was delivered (e.g., [179]).

In comparison, water losses for planets orbiting pre-main-sequence M-stars will be much higher than those for our solar system because M-star super-luminosity can last hundreds of millions of years to over 2 billion years, which can trigger runaway greenhouse conditions of comparable duration ([51]; Figures 10- 11). Depending on the model assumptions (e.g., heating efficiency, flux partitioning) and stellar EUV flux parameterization used (e.g., [186,188]), young M-star planets that later settle into the main-sequence HZ can lose up to a few tens to several *hundreds* of Earth oceans of water (e.g., [51,114,183,189–191]).

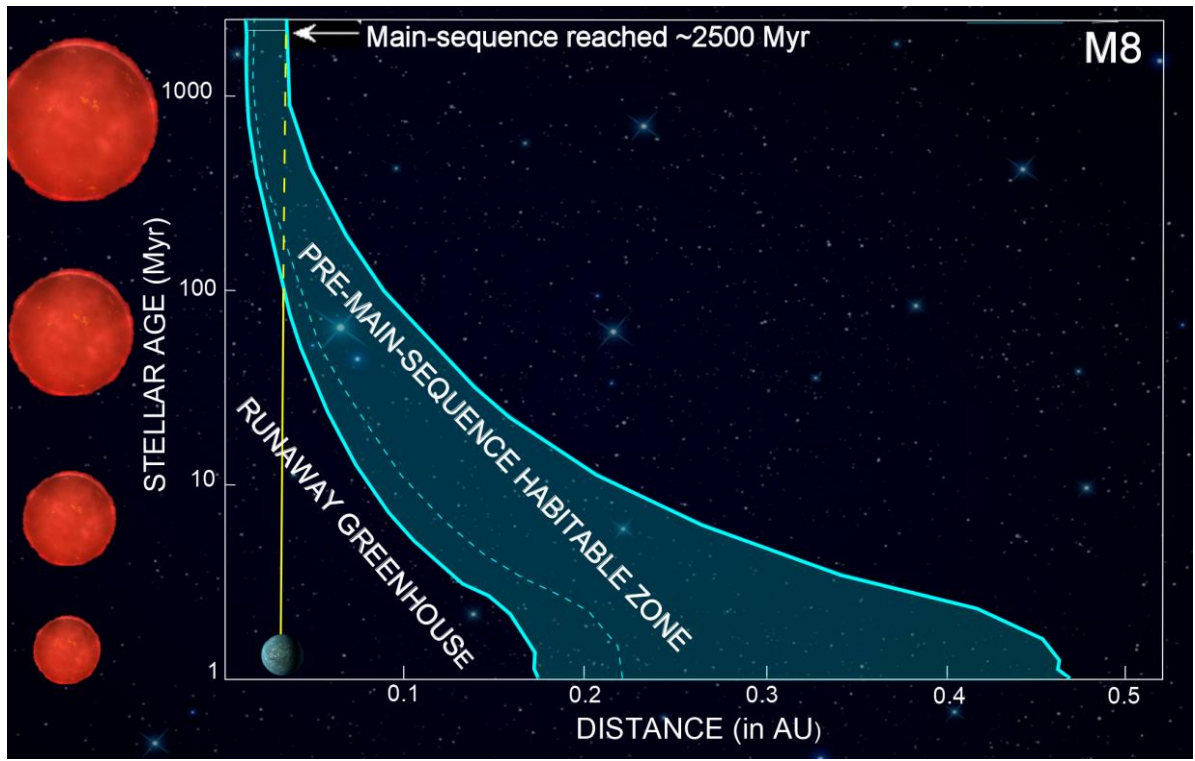

**Figure 11:** The pre-main-sequence habitable zone for an M8. A planet originally located at 0.03 AU undergoes a runaway greenhouse for ~100 Myr before settling near the outer edge of the habitable zone at the start of the main-sequence. Based on work from Ramirez and Kaltenegger [51].

*8.2 The ultimate fate of worlds during the post-main-sequence*

As main-sequence stars evolve onto the post-main-sequence, stellar luminosity increases by a few orders of magnitude and the habitable zone sweeps outward, the opposite of what occurs during the pre-main-sequence (e.g., [130,131]). For example, our Sun is expected to become up to 3,000 times brighter at the end of the red giant branch (RGB) phase as compared to today [131]. For more massive stars, the post-main-sequence luminosity increase is somewhat less, increasing ~1000x for a F1 star from the beginning of the main-sequence to the peak of the RGB (ibid).

Stellar EUV fluxes are low during the post-main-sequence so hydrodynamic escape is no longer a major concern during this time [131]. However, post-main-sequence habitability faces other serious challenges. As a star brightens to become a red giant, the star gradually loses its mass, and high stellar winds are emitted, which erode planetary atmospheres throughout the stellar system (e.g., [192,193]). In addition, the stellar density decreases, its radius grows, and to conserve angular momentum from the mass loss, the orbital radii move outward. The stellar radius increases orders of magnitude during this time. For instance, the Sun's maximum radius may extend to ~1.2 AU, more than enough to engulf the Earth according to one prediction [131], although whether the Sun will engulf the Earth is debated and depends on the stellar evolutionary model used [194].

The Ramirez and Kaltenegger [131] model for post-main-sequence A5 – M1 stars generally predicts that at least some atmosphere can be retained on large (> ~0.5 M$_{earth}$) planets or moons throughout the RGB and asymptotic giant branch (AGB) phases, so long as they are sufficiently distant (~at least beyond the Jupiter- or Saturn-equivalent distance or farther). Atmospheres on ½ Earth mass planets can survive to the end of the AGB phase for all star types at the Kuiper belt-equivalent distance. Thus, an Earth-mass exoplanetary analogue at the Pluto-equivalent distance could retain most of its atmosphere during this phase.

Post-main-sequence lifetimes can last between 200 million (A5) to 2.5 billion (K5) years for star types that can be in the post-main-sequence today (A5 – K5), assuming solar metallicity [131]. These numbers increase somewhat at higher metallicity. Although life can potentially arise and evolve during the post-main-sequence, such life need not arise during the post-main-sequence. Life could have started during the pre-main-sequence or main-sequence phases and moved to the subsurface, only unearthing during the post-main-sequence phase. For instance, if life exists in the subsurface ocean of an icy exo-Europa, it could emerge when the HZ sweeps outward during the star's RGB, melting the icy surface and exposing the subsurface ocean. The resultant atmosphere may contain potential bioindicators [131,195], with the previously-deemed subsurface life becoming surface life amenable to detection using traditional HZ criteria.

However, habitable states may not be possible on some post-main-sequence worlds orbiting hotter (G class and bluer) stars. This is because ice is more reflective on planets orbiting such stars (e.g., [196]), which requires high melting temperatures, triggering a runaway greenhouse after the deglaciation period ends [197]. However, the lower ice albedo on planets orbiting cooler (K and M) stars yields clement surface temperatures after the ice melts and the runaway is not triggered (ibid). Nevertheless, water-rich planets with small continental fractions located in the post-main-sequence HZ of bluer stars may remain habitable if they have a functional carbonate-silicate cycle [198]

It could be challenging to detect planets located within the post-main-sequence HZ because the size and flux ratios worsen as the star increases in size over time [199]. Even so, main-sequence planetary systems with planets at distances coincident with the post-main-sequence habitable zone have been detected with direct imaging, including HR 8799, Formalhaut, and Beta Pictoris [131].

After the AGB phase, the star has completely shed its outer layers and has become a dense stellar core remnant called a white dwarf, with a mass comparable to that of the Sun in a volume similar to Earth's [194]. A white dwarf HZ can be defined by assuming similar criteria for the classical HZ [200]. Fossati et al. [201] found that a planet located at ~ 0.01 AU can remain in the HZ for up to 8 billion years, which is plenty of time for complex life to develop. However, given that the white dwarf HZ is initially much farther away, such planets would have likely been in a runaway greenhouse state, which ensures desiccation unless these worlds had started out much more water-rich than the Earth [202,203]. Additionally, strong tidal forces may trigger runaway greenhouses on such close-in planets (ibid) while high impactor energies are likely to erode any atmosphere [203].

## 9. Habitability of ocean worlds

Countering the arguments in Section 8.1, recent planet formation theory suggests that M-dwarf HZ planets can potentially accrete a few tens of percent of water from the protoplanetary disk, dwarfing any losses that they would sustain during the early runaway greenhouse phase [204]. Such ocean worlds should be most common among the most massive planets (> 1 earth masses) whereas they should be less common on smaller worlds [204]. Oceans worlds may be especially prevalent in M-star systems because 1) disk densities are predicted to be higher for M-star disks (e.g., [205,206]), 2) the lack of gas giants in M-dwarf systems may facilitate volatile acquisition [207], 3) tighter orbits suggest closer ice lines [204] and 4) planets that migrate in at later times are more likely to be volatile rich, escaping the worst of pre-main-sequence losses (e.g., [178,179]). Some planets located in the pre-main-sequence HZ may fit in this latter category [51].

The recently-discovered TRAPPIST-1 system is an amazing example of a potential seven planet resonant chain, consisting of up to 5 volatile-rich planets [149,190,206,208–210]. Grimm et al. [209]

predict that TRAPPIST-1b may be a water-rich planet. However, according to its location with respect to the pre-main-sequence HZ, it would have been in a runaway greenhouse state over the entire lifetime of its star, possibly losing a few tens of Earth oceans of water in the process [191]. Thus, the only way TRAPPIST-1b (and similar planets) can still be water-rich today is if it had started extremely water-rich, possibly as an ocean world.

Nevertheless, the question remains: Are these hypothetical ocean worlds habitable? Abbot et al.[211] argue that the carbonate-silicate cycle on potentially habitable planets requires some land to operate, which suggests that ocean worlds, which have no visible land, may not be habitable. Moreover, surface pressures on ocean worlds would be high enough to shut off volcanism (e.g., [212,213]). However, Levi et al. [214] have recently developed a mechanism, originally proposed in Kaltenegger et al. [215], by which such ocean worlds, which contain a few tens of percent of their mass in water, may support life (Figure 12). The sea ice in these ocean worlds is predicted to be enriched in $CO_2$ clathrate hydrates, which keep the ocean $CO_2$-saturated. The ensuing wind-driven circulations and Eckman pump and suction mechanisms strive to degas tens of bars of $CO_2$ directly into the atmosphere. So long as subpolar (referring to latitudes between ~60 and 90 degrees) temperatures remain below freezing, sea ice enriched in clathrates can form under such high pressure atmospheric conditions. For comparison, $CO_2$ clathrate hydrates on Earth can only form within the high-pressure conditions of the sea floor. Warm (> 265 K) tropical and subtropical regions in these ocean worlds are necessary to prevent global glaciation and to bolster the circulation. Cold subpolar regions are also necessary to establish freeze-thaw cycles that help life evolve in diluted ocean worlds by concentrating the nutrients that life needs [214]. Once sea ice enriched in clathrates becomes denser than water (occurs once ice thickness exceeds ~ 1 – 2m), the ices sink, which acts to hinder global glaciation [214].

The Levi et al. [214] mechanism was recently modeled using coupled energy balance and single-column radiative convective climate models [216]. They find that high rotation rates (> ~8 hours) are necessary to generate the warm subtropical and cold subtropical temperatures required to sustain the mechanism. Thus, this mechanism is unlikely to work for M-stars cooler than about a M3, which are likely to be tidally-locked, even should they possess dense atmospheres [177]. However, assuming such ocean worlds exist, a zone equivalent to a habitable zone, can be defined for ~ G to early M-stars, in which stable climates involving sea ice enriched in clathrates can form in the subpolar regions [216]. The worlds in this ice cap zone have dense $CO_2$ atmospheres, which concentrates this region in a circular shell near the classical HZ outer edge [216]. These ocean world atmospheres can be easily distinguished from other terrestrial planets in the HZ (see Section 15.4).

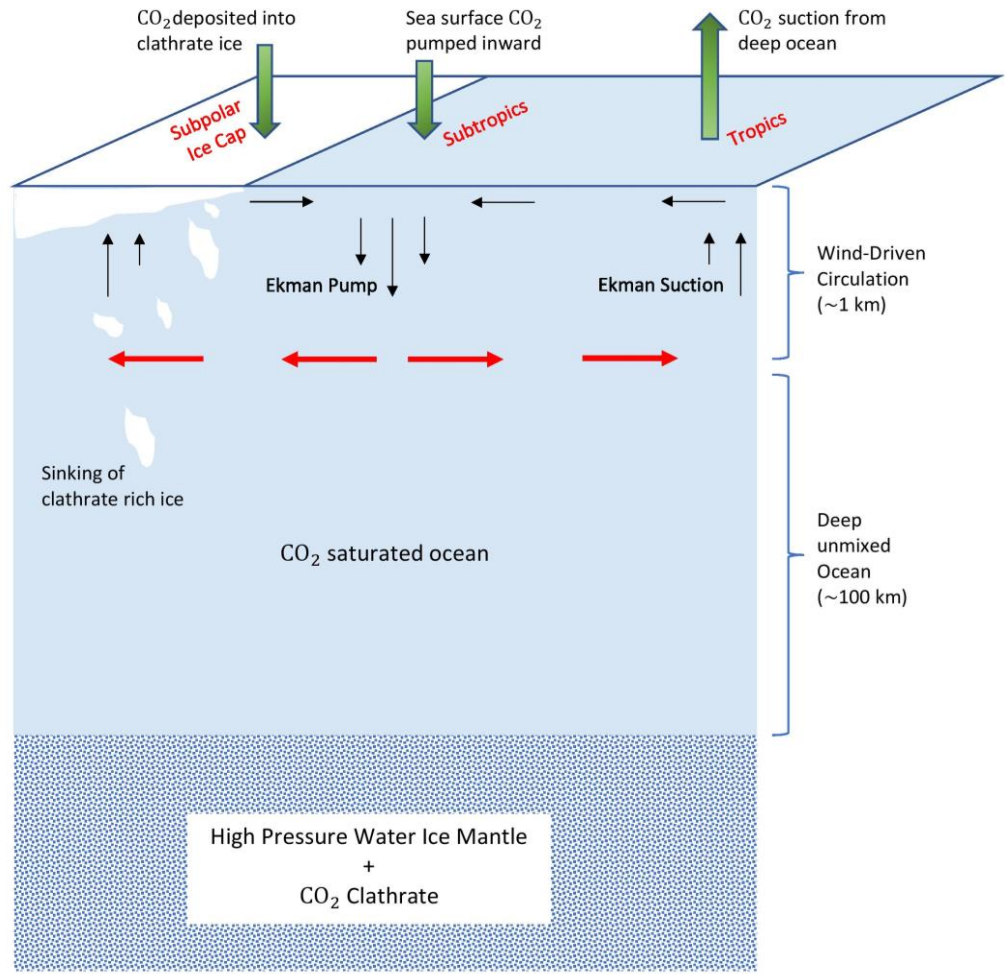

**Figure 12:** Schematic of $CO_2$ transport between ocean and atmosphere on ocean worlds. The ocean floor contains $CO_2$ clathrate hydrates while atmospheric winds drive an upper oceanic circulation that eventually degasses $CO_2$ directly into the atmosphere (adapted from Ramirez and Levi [216]).

**10. Habitability of desert worlds**

Most of this review has considered planetary habitability on worlds that are at least as wet as the Earth, if not much wetter. The surfaces of such "aqua planets" freeze if they are located too far from the star, while enhanced water vapor photolysis leading to a moist or runaway greenhouse occurs at closer distances. However, 3-D simulations find that the solar system's inner edge for "land planets" (which are worlds with an ocean inventory that is a fraction of Earth's) is ~0.77 AU [7]. This is much closer to the Sun than the 0.95 AU computed for the classical HZ (e.g., [25]), which widens the HZ overall [7]. These land planets have a reduced water vapor greenhouse effect that enhances thermal infrared emission to space and creates a dry stratosphere which limits water vapor photolysis and subsequent H escape to space [7]. Ignoring the carbonate-silicate cycle and keeping Earth's atmospheric concentration constant, Abe et al. [7] also predict that our planet would freeze at ~1.05 AU. In contrast, the smaller water inventory on the land planet weakens the ice-albedo feedback, triggering global glaciation farther away (1.14 AU) according to their model.

A subsequent calculation with a single-column radiative-convective climate model found a minimum inner edge distance of 0.38 AU for land planets if the atmospheric relative humidity is reduced to 1% [8]. However, these latter calculations may be flawed because their surfaces were not in energy balance. That is, the net absorbed solar flux (solar +thermal IR) at the surface must equal the convective (latent + sensible) heat flux [153,217]. At 1% relative humidity, Kasting et al. [153] only found balanced solutions when the mean surface temperature was ~ 250 K. Thus, solutions that are in energy balance do not exist at the high mean surface temperatures (~350 K) that Zsom et al. [8] predicted were necessary to trigger a runaway greenhouse on dry planets. Indeed, Kasting et al. [153] calculate a maximum lifetime of liquid water against evaporation for such planets of ~400 years, which implies that these worlds are likely to be dry and uninhabitable.

However, perhaps liquid water could still be present on the surface in cooler areas at higher elevation [8]. Indeed, such refuges for life may also be available on the Earth in the far distant future even as the Sun gets brighter, atmospheric $CO_2$ concentrations decrease, and C3/C4 photosynthesis ends in ~ 1 billion years [218–220]. Unicellular life will be more resistant however, lasting for 2.8 billion years from present, given the presence of sheltered, high-altitude, or high-latitude regions [220]. Thus, it is unclear whether the deserts worlds of Abe et al. [7] and Zsom et al. [8] cannot also harbor vestiges of life, at least on some regions of the planet.

Interestingly, hot desert worlds present similar observational challenges in transmission spectroscopy as do planets with Earth-like water inventories. The surfaces of hot desert worlds are still optically thick enough to hinder determinations of its surface properties (e.g., mixing ratio, temperature) [8].

**11. Binary star habitable zones**

Several studies have assessed the habitable zones of binary stellar systems following the confirmation by Kepler of their existence (e.g., [221–230]). Binary star systems are very common and comprise ~ 40- 50% of Sun-like systems [231,232] (revising the nearly 60% deduced in Duquennoy & Mayor [233], whom had overestimated their completeness correction [231]), which makes them a rich source of potentially habitable planets. There are two types of binary systems (Figure 13). In the P-type (Planet-type) or circumbinary system, the planet orbits both host stars. In contrast, the planet only orbits one of the stars in S-type (Satellite-type) systems [234].

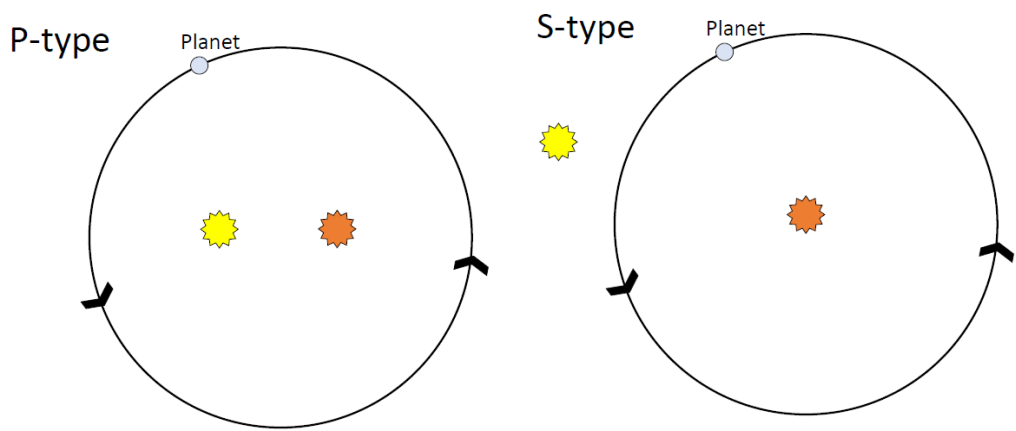

**Figure 13:** Schematic of P-type (left) and S-type (right) binary star systems.

However, modeling binary star habitable zones presents additional challenges. One of these is that stellar fluxes will vary greatly (in intensity and possibly spectral energy distribution) along the planet's orbit (e.g., [228,235,236]). The stellar fluxes received by the binary star planet are also determined via a *weighted*, not a direct, summation of the flux contributions of both stars [226–228]. This is because the heating response in atmospheres is greatly influenced by the spectral class (see Section 7.2). This has been addressed by using a representative blackbody effective temperature distribution for each of the stars [226] or via a spectral factor that explicitly weighs the stellar energy distribution (e.g., [227,228]).

P-type systems can generally be modeled as single stars by using the combined weighted fluxes of the individual hosts, assuming that the secondary star is sufficiently dim and considerably less massive than the primary [222,223,226]. However S-type systems are more difficult to model because dynamical effects can modify the HZ boundaries as a function of the binary eccentricity, semi-major axis distance, and stellar mass ratio (e.g., [224,227]). Thus, not all solutions are dynamically stable for S-type binary systems (e.g., [224,227]). Even if orbital stability is maintained, both dynamical and flux effects may cause the planet to repeatedly enter and exit the HZ.

Binary star HZ boundaries have been computed for multiple Kepler binary systems (e.g., [225–228]). Different criteria for assessing the habitability of binary star systems have been devised, which includes determining the amount of time that a planet remains within the HZ [224] to computing the fraction of habitable surface area over some time interval [229]. Subsequent work used a latitudinally-dependent energy balance model to evaluate the effect of Milankovitch cycles in binary star systems (both P- and S-type), finding that such cycles would be shorter (in some cases < 1000 years) but of comparable amplitude to similar cycles on Earth [230].

Although several papers exist on the topic, many of the factors discussed in previous sections of this review (e.g., different greenhouse gases, carbon-cycling, atmospheric escape) have yet to be considered in these calculations. Future models of binary star habitable zones that include such considerations could reveal some interesting additional insights. Furthermore, modeling the habitability of even higher order systems could also be the subject of future work.

**12. Other factors that may influence the inner and outer edges of the habitable zone**

*12.1 The effects of rotation rate*

The first set of 3-D GCM HZ calculations were performed by Yang et al. [237,238]. Whereas previous 1-D calculations had implicitly assumed that all planets were rapid rotators (e.g., [1,26,129]), Yang et al. [237,238] found that the inner edge is closer in to the star than what the classical HZ predicts if planets were slowly or synchronously-rotating. This is because enhanced dayside convection produces thick and reflective water clouds near the substellar point, reducing the net absorbed stellar flux and cooling the climate, which permits clement mean surface temperatures closer to the star [237].

Yang et al. [238] had originally found that HZ planets orbiting F – M stars can remain habitable at distances corresponding to stellar fluxes that could be more than twice (> 200%) those predicted for the classical 1-D HZ (e.g., [1,26]). However, synchronously-rotating HZ planets may only be common around late K- and M-stars [177] because HZ planets orbiting hotter stars are well outside the tidal-locking radius (e.g., [1,177]). Subsequent GCM studies, focusing on K- and M-stars, then found that the rotational and cloud effects in Yang et al. [238] had been overestimated because of orbital period scaling issues [239]. Later works with improved radiative transfer [27] and convection schemes [240] determined that rotation rates and clouds have a much smaller effect on the classical inner edge than had been previously thought. The Bin et al. [240] CAM5 simulations found that $S_{EFF}$ was only 0 – 50% larger for slow rotators, which translates to a modest ~0 – 19% decrease in inner edge distance(assuming a stellar luminosity of $0.7 L/L_{sun}$ for the $T_{EFF}$ = 4400 K K-star ) (Figure 14). However, there is reason to believe that the inner edge contrast between slowly- and rapidly-rotating planets in these simulations is still being overestimated. Previous calculations had been performed using idealized slab ocean GCMs that lack proper equator-pole ocean heat transport. In reality, ocean dynamics should reduce the day-nightside temperature contrast (e.g. [241]), as may also be expected from the Second Law of Thermodynamics. Thus, the calculation for K- M-stars should be repeated with models that dynamically couple the atmosphere and ocean. In any case, if 1-D and 3-D inner edge estimates are similar to one another for hotter stars, as recent studies suggest (e.g., [1,25,242,243]), it is probably more likely that Venus lost its water early in its history (e.g., [50]) rather than later (e.g., [238,244]).

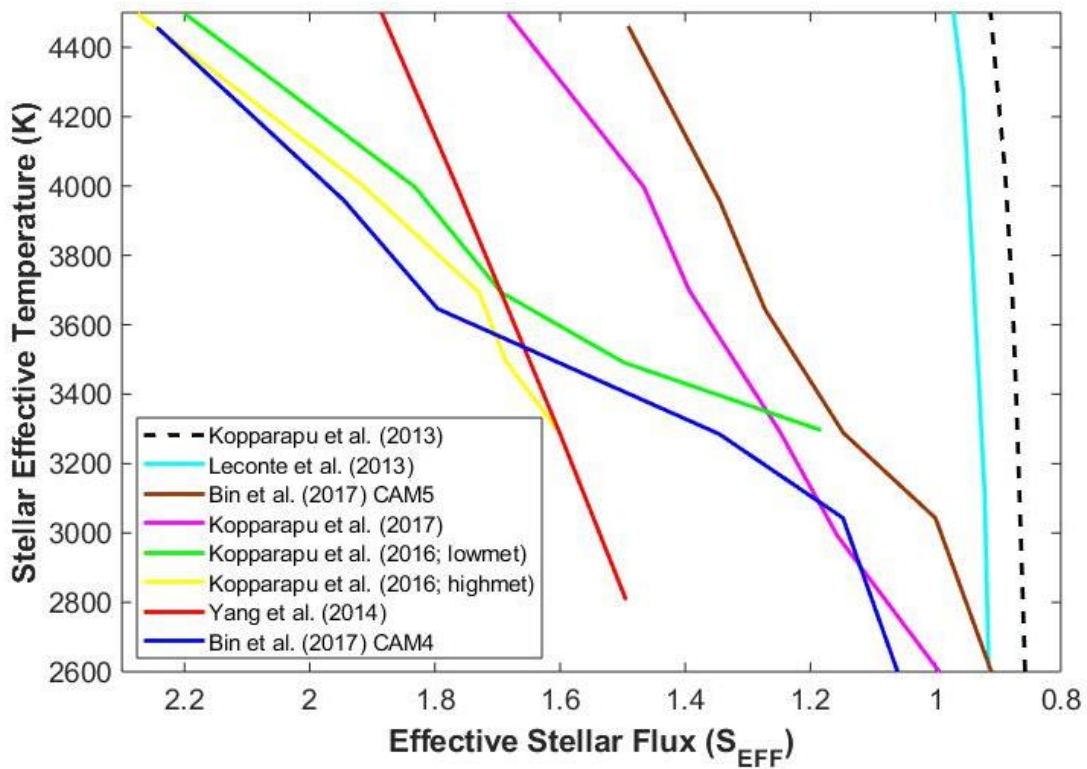

**Figure 14:** Summary of recent 1-D and 3-D model estimates of the location of the inner edge for mid-K to M-stars. The Leconte et al. [25] curve is the nominal inner edge estimate for rapidly-rotating planets, superseding the previous estimate [26]. The remaining limits are various 3-D estimates for synchronously rotating worlds.

*12.2 The effects of $CO_2$ and $CO_2$ clouds on the outer edge*

1-D modeling results suggest that 3 TRAPPIST-1 planets (e, f, and g) should be inside the classical HZ [208], which also agrees with 3-D modeling simulations [161]. Another 3-D study had initially argued that a dense $CO_2$ atmosphere cannot warm TRAPPIST-1 f and g [245], but their revised calculations now show better agreement with the other 1-D and 3-D results [246].

The 1-D and 3-D results are consistent with one another because of the high opacities that characterize atmospheres near the outer edge. Enhanced heat and dynamical transport in dense $CO_2$ atmospheres greatly reduce the day- to nightside temperature contrast (e.g., [161,176,247]), and so the single-column approximation made in 1-D models is reasonable. However, 1-D and 3-D models of tidally-locked M-star planets with thinner atmospheres exhibit greater disagreement owing to the reduced day- to night-side heat transport. Nevertheless, recent work has shown how 1-D models using the weak thermal gradient approximation can produce results for tidally-locked planets that are not only consistent with those of 3-D models, but are able to generate almost identical thermal phase curves for observational purposes (e.g., [248,249]).

Moreover, given that the 1-D models that have been used to compute HZ boundaries are unable to self-consistently compute relative humidity (e.g., [1,26,51]), such calculations typically assume a surface value of 100%, which is wetter than real atmospheres. Despite this simplification, the 1-D assumption does not significantly overestimate the outer edge distance because water vapor concentrations at the outer edge are low and water vapor absorption is minimal [216,250]. Large decreases (>10%) in the outer edge distance are predicted only at extremely low (~0%), and likely unrealistic, values of relative humidity [216].

In addition, $CO_2$ clouds should form in these atmospheres (e.g., [38]), producing significant greenhouse warming (several 10s of K) if cloud coverage approaches ~100% [38,39]. However, the warming from this mechanism is greatly reduced at lower cloud fractions. Recent 3-D simulations of early Mars predict that a typical $CO_2$ cloud coverage may be ~50%, which reduces the surface warming from $CO_2$ clouds to 15 K, even assuming idealized parameters that maximize this warming [39]. Plus, all previous studies had only used two streams in their radiative transfer calculations. After improving accuracy by increasing the number of streams, $CO_2$ clouds were found to provide almost no net greenhouse effect [40]. This is another reason why the 1-D and 3-D outer edge results appear to be consistent with one another.

**13. 1-D and 3-D Models as planetary habitability assessment tools**

Until recently, virtually all exoplanet habitability studies had been performed using single-column (1-D) radiative-convective climate models. This has changed in recent years, however, as a wide array of 1-D (including energy balance) (e.g., [2,50,51,55,132,148,216,251–255]) and 3-D (e.g., [27,161,237,239,240,244–246,256,30,257]) models are now being used to assess exoplanet habitability.

The reason for this trend is because both 1-D and 3-D climate models have complementary advantages and disadvantages. 1-D models are computationally cheap, which allows quick exploration of parameter space, the incorporation of more detailed radiative transfer, and the consideration of a larger number of greenhouse gases. 1-D models are also more easily coupled to other (e.g., interior, escape stellar, photochemical, and other 1-D) models.

In contrast, 3-D models can calculate more atmospheric processes self-consistently than can be done with 1-D models, which must make more simplifying assumptions regarding dynamics, relative humidity, and clouds. Thus, 3-D models can assess the subtleties of atmospheres in greater detail, which includes evaluating complex circulation patterns.

Thus, both 1-D and 3-D models will remain essential for studying planetary habitability, including studies of the habitable zone. One previous opinion is that 3-D models are the future of theoretical habitability studies [153,258]. However, recent years are instead unveiling a future dominated by both 1-D and 3-D models. Although 3-D modeling studies have increased, 1-D models, including coupled 1-D models, have also been on the rise. For instance, coupled 1-D atmospheric and magma ocean models can compute volatile retention and $O_2$ build up [55,116] and estimate pre-main-sequence HZ inner edge boundaries (e.g., [50,254]). Single-column radiative-convective and energy balance models are also coupled to evaluate processes in multiple dimensions (e.g., [79,216,251,252]). Coupled climate-photochemical calculations, which have been popular for many years (e.g., [170,259]), remain in high use, particularly for the calculation of atmospheric biosignatures (e.g., [255,260,261]).

Given the lack of detailed exoplanetary observations that are needed to fully utilize the potential of 3-D models, plus limitations in computer power, simpler 1-D models will remain crucial for advancing our knowledge and testing new ideas (e.g., [134,216,262]). For example, 1-D models that can treat clouds semi-consistently already exist (e.g., [263]) as do coupled radiative-convective climate and energy balance models that can emulate the dynamical behavior of idealized slab ocean GCMs (e.g., [252,264]).

However, 3-D models are also improving. For instance, once the sole domain of 1-D models, 3-D models are beginning to make preliminary biosignature predictions (e.g., [30]). Although most 3-D modeling of the inner edge of the HZ has been performed using idealized slab ocean GCMs (e.g., [27,239,240]), these lack a realistic treatment of equator to poleward heat transfer (e.g., [241]). Instead, recent 3-D calculations of the inner edge are beginning to employ fully-coupled atmospheric ocean models (e.g., [244,30]). Better cloud and convection schemes have also continued to improve such calculations (e.g., [240]).

Thus, the real future of habitability studies will be improved 1-D and 3-D models. However, better observations are truly needed to usher in a computer modeling renaissance, leading to an enhanced theoretical understanding that will advance 1-D and 3-D exoplanet habitability models alike [134]. Without better observations that can guide model predictions, all the increased model complexity, flexibility, or computational speed will be for naught.

**14. An appraisal of superhabitability**

Although the search for life has partially focused on finding a "second Earth", Heller and Armstrong [265] argue that other worlds can offer conditions that are even more suitable for life to emerge and evolve than those that have existed on our planet. These authors offer many examples of planetary traits that might characterize such "superhabitable worlds," which generally suggests that planets that are somewhat bigger, more massive, more water-rich, or dryer than the Earth (to name a few) are likely to be even more habitable than our planet. However, a growing body of work demonstrates that predicting superhabitability may be challenging and different outcomes are possible depending on the relative influences of various factors. I will highlight a few of the traits that Heller and Armstrong [265] have considered for superhabitability while giving an alternate point of view.

It was suggested that planets slightly more massive than Earth may be superhabitable because plate tectonics should last longer [265]. However, simulations indicate that basal shear stresses that are too high can induce plate failure on planets that are only slightly more massive than the Earth [64].

Although a larger water surface area (smaller continental area) was also associated with superhabitability [265], such a planet should exhibit slower continental weathering rates, allowing atmospheric $CO_2$ concentrations to rise, assuming an operative carbon-cycling mechanism [266]. This may or may not make the planet too hot for life, depending on other planetary characteristics.

Dry planets with shallower waters and more fractionate continents than the Earth may have enriched habitat diversity [265]. However, with less continental area, there should also be less plants (if otherwise an Earth-clone), and less photosynthesis and atmospheric oxygen. At low enough oxygen levels, not only would potential biosignatures be difficult to detect (e.g., [106]), but fire may not be possible [127] . Plus, the smaller volume to surface area ratio of shallower fragmented water bodies may facilitate their freezing.

Likewise, a planet with a somewhat denser atmosphere than the Earth may be deemed as superhabitable in some cases. However, it would also have a stronger greenhouse effect which may be detrimental for at least some forms of life (e.g., [217]). Depending on how warm the planet is, it could have a relatively short habitability lifetime because the runaway greenhouse threshold resulting from the slow stellar brightening would be triggered sooner (e.g., [218]).

Another argument is that HZ planets orbiting early K – late G stars (orange dwarfs) are statistically better targets to search for life, including advanced lifeforms [265,267]. This is largely based on their relatively high frequency as compared to hotter stars, modest stellar radiation environment, and larger HZ widths and orbital distances as compared to M-dwarf systems [267]. However, HZ planets around such warmer stars generally exhibit an even more clement radiation environment and wider HZs than do orange dwarfs [267], which suggests that planets orbiting stars hotter than orange dwarfs may be more superhabitable on a per-capita basis. Cuntz and Guinan [267] also argue that HZ planets orbiting stars cooler than ~a K3 are close-in enough to be susceptible to tidal-locking. However, recent work has found that planets orbiting stars as cool as an M0 (~0.5 solar masses) can avoid tidal-locking so long as planetary atmospheres are modestly thick (~1 bar) [177]. Tidal-locking can be avoided for stars as cool as an M3 (~0.3 solar masses) for planetary atmospheres that are ~10 bar thick [177]. Nevertheless, tidal-locking does not necessarily preclude habitability, although it may inhibit the formation of a planetary dynamo (e.g., [267]).

The idea that limit cycles may not occur in K-star systems is another argument that was used to support the notion of orange dwarf superhabitability [267]. This is because ice albedo for HZ planets orbiting K- and M-stars is predicted to be lower than for hotter stars (e.g., [196]), which favors non-glaciated conditions [27]. However, previous studies had employed idealized assumptions regarding silicate weathering and the nature of the ice-albedo feedback, greatly overestimating limit cycle sensitivity (see Section 4). It is therefore unclear as to whether limit cycles commonly occur on planets orbiting hotter stars either.

Thus, the relative suitability for different planets and stars to host life remains an open question. A planet's habitability is a complicated interplay of many different astronomical, atmospheric, geophysical, and (possibly) biological factors and superhabitability cannot be easily inferred. Therefore, searching for life across all relevant star types (A – M) is the best strategy. For this purpose, it does not really matter whether a planet is superhabitable or only marginally so. The focus should be in finding life in whatever form it takes and in as many places possible.

**15. Rethinking the habitable zone**

This section makes some final points about the HZ and synthesizes some of the main points made in previous sections.

*15.1. Does the habitable zone really only assess "Earth-like" life?*

One of the most important questions that mankind can attempt to answer is: "Are we alone?" In the drive to answer this age-long question, our civilization has made major progress in the last few decades, from wondering whether planets outside of our solar system really exist, to confirming the existence of nearly 4000 exoplanets (with ~4500 candidates) today [268]. As a result, the future of the search for extraterrestrial life has never been brighter. Following the success of Kepler and the recently-launched TESS (Transiting Exoplanet Survey Satellite), the (JWST) James Webb Space Telescope will soon follow suit and plans are afoot for proposed direct imaging (HabEX; Habitable Exoplanet Imaging Mission and LUVOIR; Large UV/O/IR Surveyor) and transiting (OST; Origins Space Telescope and PLATO; PLanetary Transits and Oscillations of stars) missions. Very large ground-based telescopes (TMT; Thirty Meter Telescope, GMT; Giant Magellan Telescope and ELT; Extremely Large Telescope) are also on the horizon.

However, to ensure that these missions are ultimately successful in the search for life, it is vital to employ a navigational tool that is capable of distinguishing promising targets from those that are not. For this task, the HZ currently remains far and away the best and most viable option. Despite its shortcomings and uncertain assumptions (discussed below), the classical definition is and has been in wide usage by the Kepler team (e.g., [269]) and continues to be a major influence in target selection for upcoming missions.

Nevertheless, the debate continues as to how useful the HZ really is. A common critique is that the HZ can only search for "Earth-like" life or "life as we know it" (e.g., [270,271]). It is first worth noting that "Earth-like" is a vague and commonly-used term with no consensus as to what it means nor to what degree a planet can differ from the Earth and still be considered Earth-like (alternate terms include "habitable planets" or "potentially habitable planets"). Although the classical HZ includes several assumptions that are consistent with such "Earth-like" conditions (e.g., planets orbiting main-sequence stars, the key greenhouse gases are $CO_2$ and $H_2O$), the habitable zone, even in its most restrictive classical definition, allows for a wide variety of atmospheric compositions that are strictly not like what we see on our planet. For instance, a potentially habitable planet near the inner edge of our solar system may have orders of magnitude more water vapor in its atmosphere than does the Earth. A planet near our outer edge would contain ~8 bars of atmospheric $CO_2$ (e.g., [1,26]), while possessing a drier atmosphere. Around a late M-star, enhanced absorption arising from the redshifted stellar energy distribution could allow the buildup of ~20 bars of $CO_2$ for outer edge planets (e.g., [1,26]). As also explained in Section 4, volcanic rates would need to be *many* times higher than Earth's to sustain such dense $CO_2$ atmospheres ([2,79]). The small sampling of planetary environments just mentioned exhibit conditions that are clearly unlike Earth and so assuming that any emergent life would be like ours is unsubstantiated. In fact, alien vegetation on planets orbiting other star types may photosynthesize at different wavelengths and manifest different colors than terrestrial plants ([272,273]). Previous work has modeled what the expected biosignatures over geologic history for a habitable planet might be, but these were done for an Earth clone with a 1 bar atmosphere orbiting the Sun and not for the diverse HZ planets just mentioned [108,261]. In other words, there is no a priori reason to suspect that biosignatures for many classical HZ planets would be similar to those an extraterrestrial observer might detect for the Earth. Altogether, and especially including the various different HZ formulations and recent advances discussed in previous sections (e.g., [2,8,51,129,131,132,200]), the HZ is thus equipped to assess potentially habitable planets that may *not* be "Earth-like" or "life as we know it".

*15.2. The classical HZ should be complemented with other HZ definitions*

One potential drawback of the HZ's flexibility is whether the planets it predicts actually exist. Such arguments have recently been used to cast doubt on alternative HZ definitions ([8,129]), including the existence of potentially habitable planets with dense primordial hydrogen envelopes or desert worlds[153]. However, such arguments are troublesome because the same skepticism can (and *should*) be applied to classical outer edge HZ planets with multiple bars (up to ~20) of $CO_2$ in their atmospheres. Despite the widespread usage of the classical HZ in recent decades, there is no evidence that such worlds exist either. Their existence is inferred from extrapolating the carbonate-silicate (or equivalent) cycle on Earth to suggest that atmospheres of habitable planets near the outer edge may contain many bars of atmospheric $CO_2$ (e.g., [1]). However, no *direct* observations exist of such a long-term cycle [274], as its existence has only been indirectly inferred from solubility experiments and theoretical models (e.g., [33,266,275]). Also, a multi-bar $CO_2$ atmosphere may have existed on the early Earth following accretion, but it was brief as most of that $CO_2$ was quickly subducted (within ~10 - 100 Myr) as the atmosphere cooled from an uninhabitable runaway greenhouse state [276]. Moreover, even should a universal long-term carbonate-silicate cycle operate on other habitable planets, it is not clear to what extent it does. For instance, perhaps the cycle "shuts down" beyond a certain outgassing rate or atmospheric pressure level before planets can accumulate the multi-bar $CO_2$ atmospheres that characterize the outer edge. Or, as has been argued recently [94], multi-bar $CO_2$ atmospheres closer to the outer edge would collapse under the reduced sunlight, possibly rendering such planets uninhabitable. If true, this would substantially decrease HZ width in the absence of secondary greenhouse gases ([94]). Thus, the prediction that *all* habitable outer edge planets *should* have very thick (> ~ 5 bar) $CO_2$ atmospheres is an untested assumption that needs to be verified by observations.

In contrast, indirect (although also unproven) evidence is available for terrestrial planets composed of atmospheric compositions that are consistent with alternate HZ formulations. For instance, close-in terrestrial planets with dense primordial hydrogen atmospheres likely exist, according to Kepler observations (e.g., [269]). Martian meteorites and models also favor a highly-reduced, possibly $H_2$- or $CH_4$-rich early atmosphere on early Mars ([146]), perhaps within a relatively dense, although not multi-bar, $CO_2$ atmosphere ([68,93,140,141]). This is because, the addition of secondary greenhouse gases like $CH_4$ and $H_2$, can significantly reduce the $CO_2$ pressures required to achieve warm conditions, while also providing the additional heating to counter the atmospheric collapse mentioned above ([2,132]). Thus, such secondary greenhouse gases help keep the HZ wide. Moreover, as mentioned in Section 7.1, scale heights for $CO_2$-$H_2$ atmospheres are larger than those for classical HZ dense $CO_2$ atmospheres, which facilitates the extraction of spectral information. After all, can we actually probe useful spectral information from a dense 10-bar $CO_2$ atmosphere near the outer edge? All such ideas, along with a universal carbonate-silicate cycle, should be considered as working hypotheses. Only through observations can the ideas found to be supported in nature be refined and improved for follow-up missions.

*15.3. Recommendations for using the HZ to size telescopes and find life*

Thus, the HZ size itself remains uncertain despite numerous climate modeling efforts over the years (e.g., [1,26,129,132]). An accurate estimate of HZ size is useful for determining the telescope aperture required for direct imaging. Missions use a quantity called $\eta_{earth}$, which is the estimated fraction of stars that host at least one terrestrial planet in the habitable zone (e.g., [277]). Estimates of this quantity have been made for different spectral classes (e.g., [278,279]), although large error bars exist. This is partially because observations will miss some planets, requiring statistical corrections that enable a more accurate estimate of planetary occurrence around different stars (e.g., [279,280].

If $\eta_{earth}$ is computed to be big, then potentially habitable worlds are common, which suggests that only a smaller (and cheaper) telescope is needed to find life. The opposite is true if $\eta_{earth}$ is found to be small.

Kasting et al. [153] suggest using the more pessimistic (narrower HZ) moist and maximum greenhouse limits of the classical HZ to determine the telescope diameter (or aperture). This is because $\eta_{earth}$ would be small and the telescope would then be as big as necessary according to Kasting et al. [153]. However, this assumes similar exoplanet yields in both cases. At a given telescope diameter, HZ exoplanet yields can be ~ 1.5 – 2x greater if more optimistic definitions are used (see Figure 2 in Stark et al. [135]). Exoplanet yields should be even larger than these if outer edge extensions, like those in Ramirez and Kaltenegger [132] or Pierrehumbert and Gaidos [129], are employed instead. For example, if the telescope can observe the outer edge of the $CO_2$-$H_2$ HZ (2.4 AU), the planet-star contrast ratio there would, according to the inverse-square law, only worsen by a factor of $((2.4/1.67)^2)$ two relative to the classical HZ outer edge. Designing a telescope capable of observing such distances may not be a bad idea because the $CO_2$-$H_2$ HZ includes additional terrestrial planets missed by the classical definition, like TRAPPIST-1h [149]. Striving to achieve better contrast ratios may also allow us to observe sub-Earths, like Mars, which shows abundant evidence of a once habitable planet (e.g., [68] and see Section 3.2). Such goals may justify the extra expense. Although achieving these contrast ratios near the outer edge and beyond is currently a technical challenge, it is not unreasonable to expect that technology will continue to improve in the upcoming decades.

Moreover, determining which are the "best" HZ limits to size direct imaging telescopes remains a judgement call. The angular separation between the star and the planet ($\theta$) is ~ $C\lambda/D$ ~ a/d, where $\lambda$ is the observing wavelength, D is the telescope diameter, d is the distance from Earth to the star (in parsecs), a is the planetary orbital separation (in AU), and C is a constant derived from the telescope design. For the inner working angle (e.g., inner edge), C is a smaller number whereas it is larger for the outer working angle. Therefore, the closest stars have the largest separations ($\theta$) and only smaller telescopes are needed to resolve their planetary systems (e.g. [153]). Designing a telescope around a pessimistic HZ will then require a larger aperture because $\eta_{earth}$ would be calculated to be small and the search would be confined to a smaller region of orbital space. In contrast, designing the telescope around a wider HZ implies a larger $\eta_{earth}$, which would require a smaller aperture, possibly resulting in more HZ planets , although perhaps fewer that are similar to the Earth (e.g., more desert worlds and planets located beyond the classical outer edge) [135,153]. As we learn more about the characteristics of habitable planets, proposed next generation telescopes (HabEx, LUVOIR, PLATO) could provide further information on the ideal mission architectures for subsequent missions.

However, maximizing success in the search for life will also require input from indirect detection methods, including the radial velocity and transit techniques. For example, the Kepler mission has been a resounding success, demonstrating the power of the transit technique (which finds exoplanets by measuring the slight dimming of a star as an orbiting planet passes between it and Earth) in spite of the early failure of two of its reaction wheels (e.g., [281]). The recently-launched TESS mission will also be using the transit technique to target 200,000 K- and M-stars located across the entire sky, [282]. TESS stars will be significantly brighter than those surveyed by Kepler, facilitating planetary characterization for follow-up missions, like JWST. Ground-based radial velocity measurements will continue to confirm planets initially found with the transit technique. Radial velocity observations can also find planets (by measuring the wobble in a star's orbit) that can be followed up by direct imaging.

Once the direct imaging telescope is sized, the HZ can then be used to guide search efforts [153]. However, the classical HZ is limited to finding potentially habitable planets with $CO_2$-$H_2O$ atmospheres orbiting main-sequence stars. Instead, recent HZ formulations (e.g., [2,51,129,131,132,200,216]) should be used, possibly in conjunction with the classical HZ. The newer definitions are better-equipped to find planets with both $CO_2$/$H_2O$-rich or $H_2$- or $CH_4$-rich atmospheres (e.g., [2,129,132]), potentially habitable worlds during the pre- or post-main-sequence phases of stellar evolution [51,131], possibly life-bearing ocean worlds [216], habitable worlds around A-stars [2,130–132], potential habitats in the white dwarf HZ [200], or even habitable planets orbiting binary stars (e.g., [225,229]). By observing a wide variety of plausibly habitable planets, we can maximize our chances of finding extraterrestrial life. The classical HZ should not be used as our only navigational tool.

*15.4 The HZ as a navigational filter*

Although the HZ is used to determine the region where planetary surfaces may reliably support standing bodies of liquid water, some might question the concept's utility given that life itself is not a requirement.   However, I argue that the HZ has evolved from just being a mere navigational tool for finding planets that *may* have surface liquid water to a more complete "navigational filter" capable of determining which of these are most likely to harbor life. This is for two mean reasons. First, the HZ, possibly in addition with other rubrics, can predict bulk planetary process that may be occurring or have occurred, yielding first order evaluations of planetary habitability. Secondly, the HZ is a great tool for testing scientific predictions. I explain what I mean by all of this below.

For instance, the HZ can be used to test whether HZ theory itself is valid. First, a planet near the conservative HZ inner edge should undergo enhanced water vapor photolysis in its upper atmosphere, triggering a moist greenhouse. If these calculations coincide with observations, it bolsters support for the HZ concept. Moreover, these atmospheres may also produce significant levels of abiotic oxygen [24], although a careful consideration of atmospheric composition and surface temperature conditions would be needed to infer whether life really exists or not, and if it does, what type it is (Section 5).

The classical HZ also predicts that atmospheric $CO_2$ pressures on habitable planets should increase at farther distances from the star, with maximum pressures at the outer edge (e.g., [1]). Future observations would be able to infer if such a gradient exists [274].  If it does, this would provide strong support for a long-term carbonate-silicate cycle (or equivalent carbon cycling mechanism) that operates universally on habitable planets [274]. Knowing the $CO_2$ pressure would allow estimates of the planetary volcanic outgassing rate required to support it.   Alternatively, if after observing many systems, including planets inferred to be potentially habitable (if not inhabited), no evidence of such trends is ever found, then this aspect of the HZ concept could be falsified. It is worth noting that inferring a simple $pCO_2$ gradient from inner to outer edge may be difficult in practice because real atmospheres contain secondary greenhouse gases that may confound such predictions (e.g., [2,132]). Nevertheless, observations should be able to tell if the behavior is generally correct or not.

Some models predict that some outer edge planets, or those with low volcanic outgassing rates, exhibit limit cycles [76,77,79]. If true, that would suggest that some glaciated planets near the outer edge may still be habitable upon deglaciation, although the long timescales involved may make it difficult to infer this remotely.

Although most HZ work has focused on planetary habitability during the main-sequence phase of stellar evolution (e.g., [1,26]), recent years have seen a growing appreciation of the temporal evolution of the HZ, and the pre-main-sequence in particular (e.g., [51,130]). This is because the pre-main-sequence evolution determines whether planets are still habitable during their host star's main sequence phase. This is particularly important for planets orbiting M-dwarfs (e.g., [51,114,183]). For example, although Proxima Centauri b orbits a main-sequence M-star and receives an Earth-like level of stellar insolation today, unless the planet had migrated inward later, it is likely to have been in a runaway greenhouse state for over 100 Myr ([51] and Figure 11). Depending on the size of the initial ocean inventory, this may or may not be enough to completely desiccate the planet, but the (highly likely) irradiated surface should make its potential habitability dubious [181]. Similar arguments can be made for the TRAPPIST-1 planets [208], even though 3 -4 of them may be in the current day habitable zone [149,206]. To survive, life would have to evolve extreme resistance or tolerance to such high radiation levels or find safe haven within underground or ocean habitats. None of these insights would have been possible by simply assessing main-sequence habitability. Thus, the pre-main-sequence HZ, in conjunction with the main-sequence HZ, should be used to rank which worlds are most likely to support large bodies of standing water. All else equal, a planet that was not located within the pre-main-sequence HZ, even if located in the HZ today, should be ranked lower than a planet that has always been inside the HZ.

Moreover, if dense $CO_2$-$CH_4$ atmospheres near the outer edge of hotter stars are suggestive of inhabitance [2], we should observe that a relatively large fraction of these planets are inhabited as explained in Sections 7.2- 7.3.. Supposing that an observable $pCO_2$ gradient from the inner and outer edge exists, we could infer that the outer edge distance is smaller in M-star systems that contain planets with dense $CO_2$-$CH_4$ atmospheres [2]. Also, the $H_2$ in dense $CO_2$ atmospheres is likely to be of volcanic [132] rather than primordial [129] origin. Alternatively, if $H_2$ dominates the planetary atmosphere then the source is probably primordial instead. Further, if we were to find big terrestrial planets with dense $CO_2$-$H_2$ atmospheres then we may conclude that they must have very high volcanic outgassing rates, low escape rates, or potent magnetic fields (ibid). Plus, higher scale heights for hydrogen-rich atmospheres can distinguish them from other types of rocky planets ([129,132,152]). Thus, the point here is that many planetary processes can be inferred simply by considering the implications of observations as predicted by HZ theory.

Finally, ocean worlds are predicted to require fast rotation rates in order to sustain the equator-pole temperature gradients necessary to support life [216]. Thus, planets with high rotation rates that are located near the classical outer edge should be investigated for this type of planet (ibid). Plus, the resultant dense $CO_2$ atmospheres would have low scale heights that are easily distinguished from planets with fluffier H/He envelopes. If sufficiently water-rich, these worlds would also exhibit low bulk densities that can be used to distinguish them from other types of rocky planets (ibid). An example of how different HZ definitions can be used to complement the classic one is given in Figure 15.

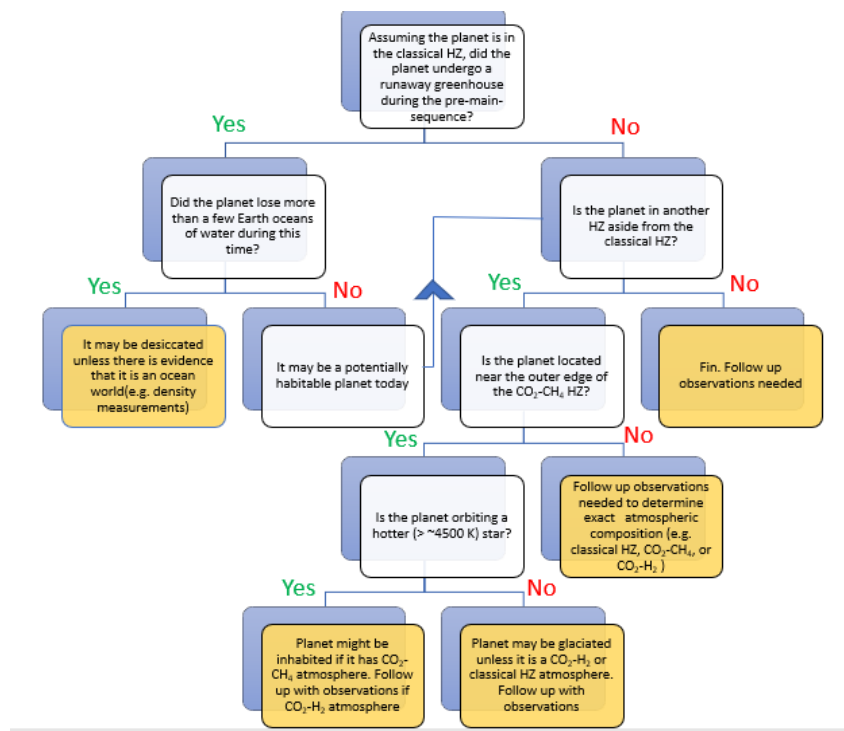

**Figure 15:** Sample flow chart using the classical HZ, along with $CO_2$-$H_2$, $CO_2$-$CH_4$, and pre-main-sequence HZ extensions to assess the potential habitability of planets. End states are in yellow.

## 16. CONCLUSIONS

To conclude, the HZ is a tool for finding potentially habitable planets. Over the years, it has evolved as an aid in finding potentially habitable (albeit exotic) planets with $CO_2$-, $CH_4$-, and even $H_2$-rich atmospheres, potentially life-bearing worlds around white dwarfs, ocean worlds, desert worlds, planets around red giant stars, worlds around pre-main-sequence stars, planets orbiting A-stars, and even worlds orbiting binary stars. The HZ can be utilized to rank potentially habitable planets in the habitable zone and is even capable of filtering out HZ worlds that are least likely to host life.

While we are not aware of another example of life besides our own, we should not pretend that we already know how life *must* be on other planets. We are not even sure how life originated on this one. Likewise, the greenhouse gas combinations and pressures that are most favorable to the emergence of life (both terrestrial and extraterrestrial) are also unknown. In other words, just because our habitable planet exhibits certain characteristics it does not follow that other habitable planets should also share those same characteristics. The principle of mediocracy is irrelevant because we have no basis to determine how common our version of life is on a universal scale (also see Appendix B2 in Heller and Armstrong [265]). We simply do not have a second example of life to compare ours with. For these reasons, it is illogical to employ any one particular version of the HZ over all other ones. Instead, we should try to find as many types of potentially habitable planets as possible, using as holistic and flexible an approach as is feasible, as has been advocated throughout this review (Section 15, in particular). Only then can we maximize our chances of success.

In contrast, an approach that is too geocentric or focused on finding "Earth-like life" may reduce such chances of success. For example, extraterrestrial observers would have deemed our own Earth uninhabitable if they simply relied on measuring our planet's oxygen levels, which were low for most of its history. Let's try to minimize such oversights by searching for life that may be unlike that which has been on Earth since the Cambrian explosion or at the Great Oxidation Event. However, this may require more capable telescopes and improved observational techniques in addition to more creative habitable zone definitions. After all, we do not know if life is rare or common in the universe. So, if we want to maximize our chances of finding life elsewhere, we should assume that life may be rare and design such missions accordingly.

To be clear, I am not advocating to minimize the search for "Earth-like life", but we should challenge ourselves with creative assessments of habitable scenarios that are unfamiliar to us. In the absence of visiting alien worlds firsthand, or a major breakthrough in direct imaging, even beyond what may be coming with next-generation missions, scientific speculation is the best path forward. For instance, although we cannot definitively determine what type of life may evolve in dense multi-bar $CO_2$ atmospheres (see Section 15.1), previous work has inferred the likely spectral signatures of Earth over time [108,261]. This is a great first step although much higher (and lower) pressures and different star types should be considered for future work. Nevertheless, assuming life exactly like Earth's can evolve elsewhere, future work could similarly assess how volcanically-outgassed $H_2$ may influence early Earth spectral signatures. Indeed, volcanically outgassed $H_2$ had been suggested as a significant constituent in the early Earth's atmosphere [46]. Alternatively, conditions dissimilar to Earth's can be assessed for likely spectral signatures, using appropriate theoretical criteria [19,152,283]. Regardless, our understanding of the expected biosignatures in atmospheres markedly different from Earth's (e.g., multi-bar $CO_2$, dense $CO_2$-$CH_4$, dense $CO_2$-$H_2$, primordial $H_2$) is very poor and more work is greatly needed in these areas.

In this review, I hope to not have only accurately surveyed the recent literature on the topic, but more importantly, I hope that the summary here could serve as a framework from which future missions could build upon. I have also strived to provide additional recommendations that can complement existing classical definitions and observational approaches. Criticisms against the HZ concept are popular (e.g., [270,271,284]), and it is this author's opinion that the underlying criticism is rooted in skepticism against classical HZ assumptions that are perceived to be too geocentric (whether they are or not) or rooted in unproven assumptions. Such critiques have some merit and partially motivate the writing of this work. Some argue that being clearer about HZ limitations using more precise language would solve some problems [271,284]. However, such criticisms may subside once we realize that life may or may not be like what we see on our own Pale Blue Dot.

**Acknowledgments:** I acknowledge funding from the Earth-Life Science Institute (ELSI). I also enjoyed discussions with Aki Roberge, Shawn Domagal-Goldman and James F. Kasting. I also thank the 4 anonymous reviewers for their many comments and suggestions. I finally thank Jacob Haqq-Misra for kindly providing Figure 4. This work also benefited from enlightening discussions about planetary habitability and the origin of life at the recent ELSI "Puzzles and Solutions in Astrobiology" workshop.

**Conflicts of Interest:** The author declares no conflict of interest.